\documentclass[journal]{IEEEtran}
\IEEEoverridecommandlockouts

\usepackage{KalmanNet}
\usepackage{mathrsfs}

\usepackage[switch,pagewise]{lineno}

\newcommand{\myVec}[1]{{\mathbf{#1}}}
\newcommand{\myMat}[1]{{\mathbf{#1}}}

\acrodef{name}[CASA-KalmanNet]{Change-Aware Self-Adaptive KalmanNet}

 \usepackage[all=normal,paragraphs=tight,floats=normal,mathspacing=normal,wordspacing=tight,
 charwidths=tight,mathdisplays=normal,leading=normal]{savetrees}

\let\oldnl\nl
\newcommand{\nonl}{\renewcommand{\nl}{\let\nl\oldnl}}

\begin{document}

\title{Change-Aware Self-Adaptive AI-Aided Kalman Filters With Neural Change Point Detection
}

\author{Wenyi Zhang, Xiaoyong Ni, Nir Shlezinger, and Zengfu Wang*
\thanks{
 W. Zhang and Z. Wang are with the School of Automation, Northwestern Polytechnical University, Xi'an 710072, China (e-mail:~zhangwenyi@mail.nwpu.edu.cn, wangzengfu@nwpu.edu.cn).
 X. Ni is with the Neuro-X Institute, \'{E}cole Polytechnique F\'{e}d\'{e}rale de Lausanne, CH-1202 Geneva, Switzerland (e-mail: xiaoyong.ni@epfl.ch).
N. Shlezinger is with the School of Electrical and Computer Engineering, Ben-Gurion University of the Negev, Be’er-Sheva 8410501, Israel (e-mail: nirshl@bgu.ac.il).
This work was supported in part by the National Natural Science Foundation of China~(Grants No.~62473317, U21B2008).
\emph{(Corresponding author: Zengfu Wang.)}
}}


\maketitle


\begin{abstract}
Reliable state estimation in dynamical systems is often challenged by model mismatches, unknown noise statistics, and temporal variations. While AI-aided Kalman filters such as KalmanNet leverage deep learning to enhance classical estimation, they remain vulnerable to distribution shifts and lack mechanisms for autonomous adaptation. This work introduces {\em \ac{name}}, an online adaptation framework that integrates a dedicated neural module, termed \emph{CPDNet}, to monitor the interpretable internal features of KalmanNet and provide soft indicators of reliability degradation. These indicators dynamically regulate an online learning process, enabling data-efficient and timely adaptation to both abrupt and gradual changes in the system without requiring additional state labels from the changed regime.
Numerical experiments on linear  and nonlinear state-space models show that \ac{name} consistently outperforms existing learning-based filters under model mismatch, while approaching the accuracy of optimal classical methods with full domain knowledge.
\end{abstract}


\acresetall

\section{Introduction}
\label{sec:intro}
\IEEEPARstart{S}{tate} estimation in dynamical systems is a fundamental task in signal processing, with a broad spectrum of applications such as tracking, navigation, {and control~\cite{bar2004estimation,102704}}. Such systems are commonly described using {\em \ac{ss}} models, which provide a probabilistic framework that captures the evolution of hidden states over time based on noisy observations~\cite{durbin2012time}. This formulation naturally leads to the \ac{kf} and its variants~\cite{gannot2008kalman}, which form one of the most widely used families of algorithms in signal processing due to their simple recursive structure, interpretability, and computational efficiency.

The classical \ac{kf} assumes that the \ac{ss} model is known and accurately described by a linear-Gaussian system~\cite{kalman1960new}. When the dynamics or the observations are nonlinear, extensions such as the \ac{ekf}~\cite{schmidt1981kalman} and the \ac{ukf}~\cite{julier2004unscented} are commonly employed. These methods still rely on precise characterization of the system model and on the assumption of Gaussian noise. However, in many practical scenarios, these models are only coarse approximations of the true underlying processes. When the actual system deviates from the assumed model, the performance of conventional filtering methods can degrade substantially.

With the growing popularity of deep learning, recent years have witnessed the emergence of \ac{ai}-aided \acp{kf}~\cite{shlezinger2025artificial}. These approaches use deep learning tools to enhance classical filtering methods by leveraging data-driven models as a form of model-based deep learning~\cite{shlezinger2020model}. Various approaches have been proposed in this domain, including hybrid methods that combine deep learning tools for system identification with classical model-based estimation~\cite{imbiriba2023augmented, masti2021learning, jouaber2021nnakf, xu2024ekfnet, COHEN2025110221}; the incorporation of learned correction terms operating in parallel to a \ac{kf}-type algorithm~\cite{satorras2019combining,welling2021hkf}; and methods that learn to parameterize the state evolution  from data~\cite{ghosh2023danse,ghosh2024deepbayes}. A notable class of \ac{ai}-aided \acp{kf}, which is particularly suitable for tracking in partially known \ac{ss} models, is KalmanNet~\cite{revach2022kalmannet} and its variants~\cite{ni2022rtsnet,choi2023split,buchnik2023latent,wang2024nonlinear, buchnik2024gsp,jung2025state}. The KalmanNet methodology recasts the \ac{ekf} as a discriminative machine learning architecture~\cite{shlezinger2022discriminative} with a \ac{dnn} that augments the Kalman gain computation, enabling robust tracking in partially known environments.

While KalmanNet has demonstrated the ability to generalize to some extent across variations in the \ac{ss} model~\cite{revach2022kalmannet}, it inherits the common vulnerability of machine learning models to distribution shifts~\cite{liu2021towards}. In practice, changes in the underlying dynamics, which can be due to abrupt structural changes or gradual drift, may degrade the learned filter. The work~\cite{ni2024adaptive} proposed to tackle this by  incorporating  hypernetworks that enable adapting to some limited level of variations without having to retrain. The work \cite{Cohen2026EMKalmanNet} combined learned Kalman filtering with the expectation maximization (EM) algorithm, focusing on smoothing tasks and assuming knowledge of which parameter changes in time, while \cite{chen2025maml} incorporated meta-learning tools to facilitate rapid retraining from supervised data. However, these works are either limited to specific variations \cite{ni2024adaptive,Cohen2026EMKalmanNet}, or require a {set of  labeled trajectories from the new regime} at inference time~\cite{chen2025maml}, an all assume knowledge on {\em when} adaptation is needed.  Consequently, existing AI-aided Kalman variants have yet to achieve the full flexibility required to operate in  time-varying SS models \emph{without online labeled data or exact model knowledge}.

Classical model-driven methods provide  tools for handling changes in \ac{ss} models when the underlying parametric structure is explicitly available~\cite{mehra1972approaches}. If the changed quantities, such as noise covariances or system matrices, are known or can be reliably estimated, they can be incorporated into the \ac{kf} recursion through covariance adaptation, parameter re-estimation, or model switching. For example, EM-based and variational Bayesian inference (VBI)-based \acp{kf} update unknown noise statistics through iterative or posterior estimation under parametric assumptions~\cite{Shumway1982,huang2021variational,bai2022novel,he2025distributed,lan2023variational}. These model-driven methods are preferable when the transition and observation functions are known, differentiable, and accompanied by suitable noise models or priors. Nevertheless, extending these principled adaptive strategies to discriminative \ac{ai}-aided \acp{kf} is non-trivial. In DNN-aided algorithms like KalmanNet, the statistical structure of the \ac{ss} model, and especially the noise distributions, is not explicitly encoded; instead, it is implicitly learned from data and embedded within the DNN weights. In this case, the changing quantity is not only an explicit covariance matrix, but also the learned filtering rule itself. Consequently, adapting the model in response to a distributional shift not observed during offline training calls for some form of {\em online learning} or {\em fine-tuning}.
Further, the demonstrated robustness of \ac{ai}-aided \acp{kf} to moderate variations suggests that not every change in the dynamics warrants adaptation, and that the extent of adaptation required is dependent on the nature and severity of the variation. This observation motivates developing an online adaptation method tailored for \ac{ai}-aided \acp{kf}, which facilitates rapid and data-efficient learning-based adaptation with limited computational overhead and without  additional labeled data.

In this work, we introduce a novel
{online test-time adaptation framework} termed {\em \ac{name}}, designed to enable such \ac{ai}-aided \acp{kf} to operate effectively in time-varying dynamical environments. Our approach addresses the challenge of adapting discriminative neural architectures, and particularly KalmanNet, to shifts in the underlying \ac{ss} model without requiring explicit knowledge of the changing system parameters or access to labeled data.
Therefore, \ac{name} is complementary to model-driven adaptive filters: model-driven methods remain natural choices under accurate parametric knowledge, whereas the proposed method targets partially known and neural-filter-based settings where reliability monitoring and label-free online correction are required.
In particular, drawing inspiration from recent advances in asynchronous online learning for wireless communications~\cite{uzlaner2025asynchronous}, our key insight is that effective adaptation need not rely on detecting changes in the dynamics themselves, but rather on monitoring the reliability of the filter. Specifically, we propose to observe the degradation in performance of the \ac{ai}-aided \ac{kf} as an {\em implicit indicator} of a distributional mismatch, thereby enabling data-efficient and computationally lightweight adaptation.

To realize this, \ac{name} incorporates two core innovations. First, we design an independent \ac{dnn}-based module, termed \textit{CPDNet}, which continuously monitors KalmanNet during operation and generates an online measure indicating whether the filter remains valid for the current dynamical regime. CPDNet particularly leverages the interpretable \ac{ekf}-based architecture of KalmanNet,
{utilizing its internal prediction of the observation for obtaining
features that are informative on potential degradation without requiring
ground-truth state labels.}
Second, we integrate this indicator into a tailored
{online adaptation mechanism}
mechanism, which is selectively activated based on the output of CPDNet. Once triggered, this adaptation mechanism leverages the magnitude of the detected change to guide the intensity (and consequently, the duration) of the adaptation process. This  enables rapid, reliable, and {ground-truth-free} adaptation of \ac{ai}-aided \acp{kf} to changing environments with limited computational burden.

Our main contributions are summarized as follows:
\begin{itemize}
    \item
    We formulate a change-aware online test-time adaptation framework for \ac{ai}-aided \acp{kf} under partial model knowledge and nonstationary noise statistics.

    \item
    We develop \textit{CPDNet}, an independent reliability-monitoring network that exploits the interpretable innovation-related internal features of KalmanNet to estimate whether the deployed filter has become misaligned with the current dynamics. CPDNet does not modify the KalmanNet gain architecture; instead, it provides an additional self-diagnostic capability that standard KalmanNet lacks.

    \item
    We propose a closed-loop change-aware online adaptation mechanism in which the CPDNet score gates and scales a KalmanNet update based on the innovation loss, without requiring additional state labels from the changed regime.
    Unlike continuously updated unsupervised KalmanNet, which may update indiscriminately and over-adapt, the proposed mechanism selectively regulates adaptation in response to detected changes.

    \item
    We validate the proposed framework under abrupt and gradual changes in noise statistics and model parameters, showing that the proposed method improves robustness over KalmanNet and its unsupervised online-update counterpart while approaching model-aware classical filters when full post-change information is available. We further validate the method on a real robot trajectory using the NCLT dataset~\cite{ncarlevaris-2015a} under controlled observation-noise nonstationarity, showing improved full-state tracking among learned baselines without access to the covariance change, with substantially fewer online updates.

\end{itemize}

The rest of this paper is organized as follows: Section~\ref{sec:System Model and Preliminaries} briefly reviews \ac{ss} models and the \ac{ekf}.  CPDNet is detailed in {Section~\ref{sec: Method}, and is incorporated into the \ac{name} online self-adaptation framework in Section~\ref{sec:Continual}.} Section~\ref{sec: Numerical Study} details our numerical study, and Section~\ref{sec:conclusion} provides concluding remarks.

Throughout this paper, we use  boldface lowercase for vectors, e.g., $\gvec{x}$, and boldface uppercase letters, e.g., $\pmb{M}$, for matrices. The $i$th element of some vector $\gvec{x}$ is denoted $[\gvec{x}]_i$. Similarly, $[\pmb{M}]_{i,j}$ is the $(i,j)$th element of a matrix $\pmb{M}$.
We use  $\mathbb{R}$ for the set of real numbers. The operations $(\cdot)^\top$ and  $\| \cdot \|$  are used for transpose and  $\ell_2$ norm,  respectively.


\section{System Model and Preliminaries}
\label{sec:System Model and Preliminaries}
In this section we introduce the system model and recall some  preliminaries. We first present the \ac{ss} model in Subsection~\ref{subsec: Signal Model} and formulate the state estimation problem in Subsection~\ref{ssec:Problem}. Then, we review necessary preliminaries in model-based and \ac{ai}-aided Kalman filtering in Subsection~\ref{ssec:Preliminaries}.

\subsection{System Model}
\label{subsec: Signal Model}
We consider a general discrete-time dynamical system described by a time-varying nonlinear \ac{ss} model:
\begin{subequations}\label{eq:ss}
\begin{align}
    \mathbf{x}_t &= f(\mathbf{x}_{t-1}) + \mathbf{w}_t, & \mathbf{w}_t &\sim P_t(\mathbf{w}), & \mathbf{x}_t &\in \mathbb{R}^m, \label{eq:ss_state} \\
    \mathbf{y}_t &= h(\mathbf{x}_t) + \mathbf{v}_t, & \mathbf{v}_t &\sim P_t(\mathbf{v}), & \mathbf{y}_t &\in \mathbb{R}^n. \label{eq:ss_obs}
\end{align}
\end{subequations}
In \eqref{eq:ss}, $\mathbf{x}_t$ denotes the latent state vector at time $t$, and $\mathbf{y}_t$ is the corresponding observed measurement. The functions $f(\cdot)$ and $h(\cdot)$ describe the (possibly nonlinear) state transition dynamics and the observation model, respectively. The process noise $\mathbf{w}_t$ and the observation noise $\mathbf{v}_t$ are drawn from temporally independent distributions, denoted $P_t(\mathbf{w})$ and $ P_t(\mathbf{v})$, respectively, that  may vary over time to capture non-stationarities.

A widely used special case of~\eqref{eq:ss} is the \emph{linear Gaussian \ac{ss} model}, where  both $f(\cdot)$ and $h(\cdot)$ are linear functions, represented by the matrices $\mathbf{F} \in \mathbb{R}^{m \times m}$, and $\mathbf{H} \in \mathbb{R}^{n \times m}$, and the noises are Gaussian. In this case, \eqref{eq:ss} becomes
\begin{subequations}\label{eq:ss_linear}
\begin{align}
    \mathbf{x}_t &= \mathbf{F} \mathbf{x}_{t-1} + \mathbf{w}_t, \\
    \mathbf{y}_t &= \mathbf{H} \mathbf{x}_t + \mathbf{v}_t.
\end{align}
\end{subequations}
 Gaussianity implies that $P_t(\mathbf{w})$ and $ P_t(\mathbf{v})$ are zero-mean Gaussian distributions with time-varying covariance matrices $\mathbf{Q}_t$ and $\mathbf{R}_t$, respectively.

The general model in~\eqref{eq:ss} encompasses a broad class of dynamical systems. In this work, we allow the system to exhibit both nonlinear dynamics and non-stationary noise statistics, which presents a significant challenge for reliable and efficient state estimation, as detailed next.

\subsection{Problem Formulation}
\label{ssec:Problem}

We consider the filtering task (i.e., online state estimation) in a dynamical system governed by the \ac{ss} model of~\eqref{eq:ss}. Given a sequence of noisy observations $\{\mathbf{y}_\tau\}_{\tau=1}^t$ up to time $t$ and knowledge of the initial state $\mathbf{x}_0$, the goal is to generate an accurate estimate of the latent state $\mathbf{x}_t$ at each time step. Formally, the filtering problem aims to compute a mapping
\begin{equation}
\label{eqn:filtering}
    \hat{\mathbf{x}}_t = \mathcal{F}_t\left(\{\mathbf{y}_\tau\}_{\tau=1}^{t}\right),
\end{equation}
where $\mathcal{F}_t(\cdot)$ denotes the filtering rule at time $t$, which outputs the estimate $\hat{\mathbf{x}}_t$ of the true latent state $\mathbf{x}_t$.

We  focus on scenarios where  the filtering task is challenged by various sources of uncertainty and mismatch between the assumed model and the true system behavior. In particular, we identify the following key challenges:
\begin{enumerate}[label=\textbf{C\arabic*}, leftmargin=*]
    \item \label{itm:Approx} \textbf{Model Approximation:} The state evolution function $f(\cdot)$ and the observation function $h(\cdot)$ are often only coarse approximations of the true underlying dynamics.

    \item \label{itm:Noise} \textbf{Unknown Noise Distributions:} The statistical properties of the process noise $\mathbf{w}_t$ and observation noise $\mathbf{v}_t$, i.e., $P_t(\mathbf{w})$ and $P_t(\mathbf{v})$, are unknown, and may deviate from the Gaussian assumptions typically made in classical filtering.

    \item \label{itm:varying} \textbf{Non-Stationarity:} The noise distributions $P_t(\mathbf{w}), P_t(\mathbf{v})$ may change over time without prior knowledge of when or how these changes occur.
\end{enumerate}

To facilitate the design of filtering mechanisms that can cope with \ref{itm:Approx}-\ref{itm:varying}, we assume access to a labeled offline dataset $\mathcal{D}$ consisting of input-output pairs, i.e.,
\begin{equation}
\label{eqn:data}
    \mathcal{D} = \{(\{\mathbf{y}_\tau^{(i)}, \mathbf{x}^{(i)}_\tau\}_{\tau=1}^{T})\}_{i=1}^{|\mathcal{D}|},
\end{equation}
generated under a known time horizon $T$. This dataset is used for learning purposes  prior to deployment. The offline data is assumed to be decomposable into two subsets:
\begin{itemize}
    \item A {\em stationary} set  $\mathcal{D}_{\rm s}$, originating from a single system regime, in which the noise signals follow (unknown) stationary distributions denoted by $P_0(\mathbf{w})$ and $P_0(\mathbf{v})$.
    \item A {\em nonstationary} set $\mathcal{D}_{\rm v}$, where in each input-output trajectory the distribution of the noises can change in an unknown manner.
\end{itemize}
During inference, the system may encounter different nonstationary dynamics, but no labeled data or prior knowledge of the  change point from these  distributions are assumed to be available.

Specifically, in our numerical experiments (in Section~\ref{sec: Numerical Study}), a separate nonstationary dataset $\mathcal{D}_{\rm v}$ is generated for each type of \ac{ss} model variation considered, such as changes in the noise statistics or selected \ac{ss} model parameters. The training and testing trajectories are disjoint. They share the same type of parameter variation, but use independently generated trajectories and different change times.

\subsection{Preliminaries}
\label{ssec:Preliminaries}

\subsubsection{Model-Based Kalman Filtering}
\label{sssec:KalmanFilter}
When the \ac{ss} model is fully known and the noises are Gaussian, state estimation can be reliably performed using model-based filtering algorithms based on variants of the \ac{kf} for nonlinear dynamics, such as the \ac{ekf}. Kalman-type filters  track the first and second-order moments of the state, which at time $t$ are respectively denoted $\hat{\mathbf{x}}_{t}$ and $ \boldsymbol{\Sigma}_{t}$,  via recursive prediction and update steps.

Specifically, at each time $t$, the \ac{ekf} propagates the first-order moments of the state and observations via
\begin{subequations}
\begin{align}
\label{eqn:predict}
    \hat{\mathbf{x}}_{t|t-1} &= f(\hat{\mathbf{x}}_{t-1}), \\
    \hat{\mathbf{y}}_{t|t-1} &= h(\hat{\mathbf{x}}_{t|t-1}).
\end{align}
\end{subequations}
This is followed by the computation of the \ac{kg} based on linearizations of the transition and observation models as $\mathbf{F}_t = \left. \frac{\partial f}{\partial \mathbf{x}} \right|_{\hat{\mathbf{x}}_{t-1}}$ and $\mathbf{H}_t = \left. \frac{\partial h}{\partial \mathbf{x}} \right|_{\hat{\mathbf{x}}_{t|t-1}}$,  such that
\begin{subequations}
\begin{align}
\label{eqn:kalmangain}
    \boldsymbol{\Sigma}_{t|t-1} &= \mathbf{F}_t \boldsymbol{\Sigma}_{t-1} \mathbf{F}_t^\top + \mathbf{Q}_t, \\
    \mathbf{S}_t &= \mathbf{H}_t \boldsymbol{\Sigma}_{t|t-1} \mathbf{H}_t^\top + \mathbf{R}_t, \\
    \mathbf{K}_t &= \boldsymbol{\Sigma}_{t|t-1} \mathbf{H}_t^\top \mathbf{S}_t^{-1}.
\end{align}
\end{subequations}
The state estimate and its error covariance are  updated as
\begin{subequations}
\label{eqn:update}
\begin{align}
    \hat{\mathbf{x}}_{t} &= \hat{\mathbf{x}}_{t|t-1} + \mathbf{K}_t \Delta \mathbf{y}_t, \label{eqn:updateS} \\
    \boldsymbol{\Sigma}_{t} &= (\mathbf{I} - \mathbf{K}_t \mathbf{H}_t) \boldsymbol{\Sigma}_{t|t-1},\label{eqn:updateC}
\end{align}
\end{subequations}
where $ \Delta \mathbf{y}_t = \mathbf{y}_t - \hat{\mathbf{y}}_{t|t-1}$ is the {\em innovation process}.

Alternative variants of the \ac{kf} for nonlinear dynamics, such as the \ac{ukf}~\cite{julier2004unscented} and the Cubature \ac{kf}~\cite{arasaratnam2009cubature}, differ in how they approximate the propagation of means and covariances in nonlinear models, but all rely on a faithful and complete specification of the model dynamics and noise distributions. Consequently,  model-based filters are limited when facing model mismatches and uncharacterized noise, i.e., under challenges \ref{itm:Approx} and \ref{itm:Noise}.

\subsubsection{AI-Aided Kalman Filtering}
\label{sssec:KNet}
To address the limitations of model-based filtering under partial or mismatched knowledge, \ac{ai}-aided Kalman filters integrate deep learning with classical state estimation principles. A representative architecture in this family, that is the focus of the current work, is \textit{KalmanNet}~\cite{revach2022kalmannet}, which adopts a model-based deep learning approach~\cite{shlezinger2023model} by embedding the structure of the \ac{ekf} with a trainable \ac{dnn}. 

KalmanNet operates by replacing the \ac{kg} computation of the \ac{ekf} with a \ac{dnn}, typically an \ac{rnn}~\cite{revach2022kalmannet} or a connection of \acp{rnn}~\cite{choi2023split}, with alternative architectures using attention mechanisms~\cite{wang2024nonlinear}. Letting $\dnnParams$ denote the \ac{dnn} parameters, at each time step $t$, the architecture receives as input engineered features derived from the filtering pipeline, such as the predicted innovations and changes in observations, and outputs a learned \ac{kg} $\hat{\mathbf{K}}_t(\dnnParams)$. The learned \ac{kg} is then used in the classical correction step. The resulting mapping of \eqref{eqn:filtering} becomes
\begin{equation}
    \hat{\mathbf{x}}_{t}(\mathbf{y}_t; \dnnParams) = \hat{\mathbf{x}}_{t|t-1} + \hat{\mathbf{K}}_t(\dnnParams) \Delta \mathbf{y}_t .
\end{equation}

Training is performed in a supervised manner using a dataset $\mathcal{D}$ consisting of observation sequences and their corresponding ground-truth state trajectories as in \eqref{eqn:data}.
The \ac{dnn} parameters are typically learned by minimizing the empirical \ac{mse} between the true and estimated states,
\begin{equation}
    \mathcal{L}_{\mathcal{D}}^{\rm MSE}(\dnnParams) = \frac{1}{|\mathcal{D}|T} \sum_{i=1}^{|\mathcal{D}|} \sum_{t=1}^T \left\| \mathbf{x}_t^{(i)} - \hat{\mathbf{x}}_t(\mathbf{y}_t^{(i)}; \dnnParams) \right\|^2.
    \label{eqn:LossMSE}
\end{equation}

This discriminative training paradigm enables KalmanNet to learn from data how to  correct state estimates, without explicit modeling of the noise distributions or exact system dynamics. As a result, KalmanNet is well-suited for scenarios where only partial domain knowledge is available, successfully addressing \ref{itm:Approx} and \ref{itm:Noise}.
However, the learned model remains specialized to the distributional regime encountered during training. While empirical results demonstrate that KalmanNet can generalize to moderate variations in the dynamics~\cite{revach2022kalmannet}, substantial changes in the underlying distributions necessitate adaptation. This is because KalmanNet does not include mechanisms for online adaptation, and thus cannot autonomously cope with  \ref{itm:varying}.

\section{Neural Change Point Detection}
\label{sec: Method}
This section introduces a key ingredient of our online adaptation framework, which is the \ac{dnn}-based algorithm for monitoring deviations in \ac{ai}-aided \acp{kf}. We begin by explaining our design rationale in Subsection~\ref{ssec:rationale}. Then, we present the \ac{dnn}-based \ac{cpd} algorithm termed {\em CPDNet} in Subsection~\ref{subsec: Algorithm}, and explain how it is trained in Subsection~\ref{subsec: Training}.

\subsection{Rationale}
\label{ssec:rationale}
Our first step toward addressing the core challenges outlined in \ref{itm:Approx}-\ref{itm:varying} is the integration of \ac{cpd} mechanisms into \ac{ai}-aided \acp{kf}. While \ac{cpd} traditionally refers to the task of identifying abrupt changes in the statistical properties of a sequence~\cite{aminikhanghahi2017survey}, applying conventional \ac{cpd} methods directly to our setting does not leverage a critical property of KalmanNet: its ability to generalize across moderate variations in the system dynamics~\cite{revach2022kalmannet}. As a result, standard \ac{cpd} approaches, which rely on detecting shifts in the data distribution (e.g., via statistical divergences or kernel-based techniques~\cite{adams2007bayesian,alami2020restarted,harchaoui2008kernel,song2024kernal}), may trigger unnecessary adaptation events and do not distinguish benign fluctuations from distributional shifts that truly compromise filter reliability.

An alternative perspective emerges from the machine learning paradigm of concept drift~\cite{lu2018learning}. These works propose tools to assess whether a learned model remains suitable over time, and include methods such as the Drift Detection Method of~\cite{gama2004learning} and model-based statistical tests like Hotelling’s test~\cite{uzlaner2024concept}. However, these approaches are typically designed for performance evaluation rather than real-time decision-making. Moreover, they often assume access to ground-truth labels or performance feedback, which are not available in our setting, as state trajectories during inference remain unknown.

Nevertheless, a unique opportunity for unlabeled \ac{cpd} arises from the structure of KalmanNet itself. While KalmanNet is a hybrid-driven architecture, it inherits its formulation from the \ac{ekf}, 
%
whose internal features (e.g., the innovation process) implicitly encode information related to the  confidence of the filter and its alignment with the true dynamics. Consequently,  by monitoring such internal features and outputs, one can potentially detect when KalmanNet begins to diverge from reliable operation. However, unlike the model-based setting, where the relationship between these features and model validity is tractable and analytically understood, the mapping in KalmanNet is far more intricate and entangled with the learned network weights. This motivates the use of  deep learning architectures to capture and learn the complex relationships between observed behavior and the need for adaptation.

\subsection{Change Point Detection Algorithm}
\label{subsec: Algorithm}
Based on the above rationale, we design \textit{CPDNet}, a neural \ac{cpd} module specifically designed to operate alongside KalmanNet. CPDNet monitors the reliability of KalmanNet by detecting the need for model adaptation. It is formulated as a neural module with parameters $\myVec{\phi}$ that receives at each time instance $t$ a sequence of processed filtering features, denoted  $\myVec{z}_{t}$,  and produces a soft prediction $\hat{p}_{\myVec{\phi}}(\myVec{z}_{t}) \in [0, 1]$ representing a reliability-degradation score indicating whether the filter is no longer aligned with the current dynamics.
A high value  serves as a data-driven trigger for initiating adaptation in the filtering model.
Specifically, $\hat{p}_\phi(\mathbf{z}_t)$ is a data-driven reliability score
calibrated into $[0,1]$ via the sigmoid-tanh mapping and the training procedure (detailed in Subsection~\ref{subsec: Training}).
While it behaves analogously to a probability measure, it serves as
a heuristic indicator of filter misalignment rather than a strict
Kolmogorov probability.

\begin{figure}
    \centering
    \includegraphics[width=\linewidth]{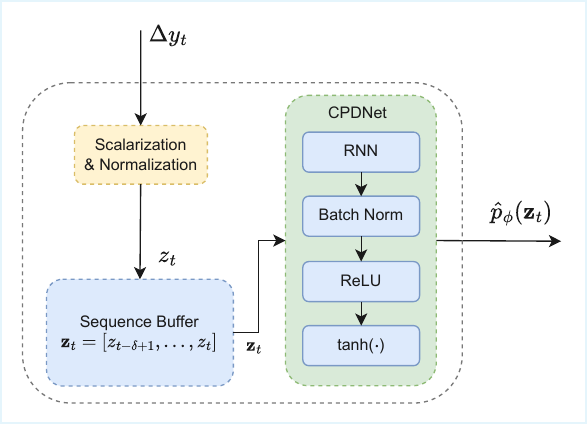}
    \caption{{Overall architecture of  CPDNet framework. KalmanNet's predict step   computes the innovation $\Delta \mathbf{y}_t$, which is then passed to the Data Process module. This module applies a nonlinear transformation to generate the normalized feature $\myVec{z}_t$, which is stored in the sequence buffer and subsequently used by CPDNet. }
}
    \label{fig:CPDNet}
\end{figure}

We next present CPDNet by detailing the construction of its input features $\myVec{z}_{t}$ and its neural architecture.  The overall procedure is illustrated in Fig.~\ref{fig:CPDNet}.

\subsubsection{Input Feature Design}
Based on the rationale outlined in Subsection~\ref{ssec:rationale}, we use the innovation process $\Delta\mathbf{y}_t$ as indicative features for detecting variations. To facilitate learning, we further compress these multivariate quantities into scalar features in the range $[0,1]$. This is achieved through handcrafted extraction of the form
\begin{equation}
    z_t = {\rm tanh}\left(\gamma \cdot \left(\sigma\left(\|\Delta\mathbf{y}_t\|\right) -0.5 \right)\right),
    \label{eqn:FeatureExt}
\end{equation}
where $\sigma(\cdot)$ is the sigmoid function and $\gamma$ is a  hyperparameter.

The feature construction in \eqref{eqn:FeatureExt} enables a decoupled, sequential non-linear mapping: the initial $\sigma(\|\Delta \mathbf{y}_t\|) - 0.5$ transforms the $\ell_2$ norm into a normalized "influence" score within $[0, 0.5)$. This first stage specifically captures the non-linear  behavior of the innovation process magnitude, while establishing a zero influence for zero input. Subsequently, the $\tanh(\cdot)$ function operates on this pre-processed "influence" (scaled by $\gamma$), performing a second-stage non-linear compression to a bounded score range. This two-step process allows the model to capture distinct non-linearities in mapping the innovation magnitude into an influence score, which is then used as the input feature for CPDNet.
To reduce sensitivity to isolated anomalies, we aggregate the transformed features over a sliding window of length $\delta \geq 1$, such that the features processed by the CPDNet are given by:
\begin{equation}
\label{eqn:features}
    \myVec{z}_{t} = [z_{t-\delta+1}, \ldots, z_{t}].
\end{equation}
The window length $\delta$ determines the temporal receptive field of CPDNet. It should be selected according to the expected time scale of the nonstationarity: a larger $\delta$ improves robustness to isolated fluctuations but may delay the response to abrupt changes, while a smaller $\delta$ improves responsiveness but may be more sensitive to noise. If the underlying drift is much slower than $\delta$ and only induces very weak local changes in the innovation statistics, CPDNet may delay or miss the adaptation trigger. Thus, $\delta$ should be interpreted as a detection-time-scale hyperparameter rather than a universal guarantee for all drift rates.

\subsubsection{Neural Architecture}
 CPDNet is structured to be able to learn to exploit temporal dependencies  between the entries of the input stream $\myVec{z}_t$, with an output layer designed to produce a non-negative bounded reliability score. As such, it consists of an interconnection of an \ac{rnn}, applied to the windowed time sequence $\myVec{z}_t$ in a many-to-one fashion, and a fully-connected layer that produces scalar outputs, with an output ReLU activation followed by a scaled $\tanh$ to ensure bounded score outputs in $[0,1]$. The ReLU activation suppresses negative pre-activation values and therefore biases weak evidence of model mismatch toward a zero CPDNet score, while the final decision of whether to update is explicitly controlled by the thresholding rule in \eqref{eqn:TVLearningRate}.
This design leverages the sequential modeling capabilities of \acp{rnn} while ensuring that the output is well-calibrated and robust to fluctuations.

\subsection{Training CPDNet}
\label{subsec: Training}
CPDNet is designed to enable online adaptation of KalmanNet. However, while the adaptive processing of KalmanNet that uses CPDNet (detailed in the following section) is done online, CPDNet itself is trained in advance using the data in \eqref{eqn:data} and by formulating a supervised loss measure. This training procedure is comprised of two stages: $(i)$ Stationary reference  pre-training; and $(ii)$ Nonstationary change-detect training. We next elaborate on these stages.


\subsubsection{Stationary Reference Pre-Training}
The first stage involves training a reference KalmanNet model using a stationary dataset $\mathcal{D}_s$.
The purpose of introducing this auxiliary \ac{dnn} is to facilitate both the generation of training data and the construction of a supervised-style loss function, which can then be leveraged to train and evaluate CPDNet. This design choice is inspired by strategies commonly adopted in adversarial learning~\cite{miller2020adversarial}, where an additional network is employed to provide informative feedback signals during model optimization.

Letting $\myVec{\theta}$ denote the parameters of the reference KalmanNet, these parameters are trained guided by the standard \ac{mse} loss of \eqref{eqn:LossMSE} based on the stationary $\mathcal{D}_s$, i.e., this learning stage aims at recovering
\begin{equation}
    \myVec{\theta}^\star = \mathop{\arg \min}_{\myVec{\theta}} \mathscr{L}_{\mySet{D}_{\rm s}}^{\rm MSE}(\myVec{\theta}).
\end{equation}

\subsubsection{Nonstationary Change-Detect Training}
Next, we fix the reference KalmanNet, and train CPDNet to monitor its behavior when applied to the nonstationary data $\mySet{D}_{\rm v}$. Since this dataset contains both observations and state trajectories, we can evaluate the accuracy of KalmanNet, and in turn  evaluate CPDNet during training based on its ability to identify when KalmanNet becomes inaccurate.

 To convert the error of KalmanNet, i.e., $\big\| \mathbf{x}_t^{(i)} - \hat{\mathbf{x}}_t(\mathbf{y}_t^{(i)}; \dnnParams) \big\|$, into a bounded degradation target comparable to the CPDNet output, we repeat the feature conversion used in \eqref{eqn:FeatureExt}, and set
\begin{equation}
        p_t^{(i)}(\dnnParams) \!=\! {\rm tanh}\left(\gamma  \left(\sigma\left(\| \mathbf{x}_t^{(i)} - \hat{\mathbf{x}}_t(\mathbf{y}_t^{(i)}; \dnnParams)\|\right)\! - \! 0.5 \right)\right).
    \label{eqn:Label}
\end{equation}
The same $\gamma$ is used in \eqref{eqn:FeatureExt} and \eqref{eqn:Label} to consistently scale both mappings into $[0,1]$.
The probabilistic quantities in \eqref{eqn:Label} approach one when the $\ell_2$ error of KalmanNet is large, and are thus used as labels for signaling that adaptation is required. Since  $\mySet{D}_{\rm v}$ contains time-varying noise distributions, the application of KalmanNet trained on the stationary data $\mySet{D}_{\rm s}$ is expected to lead to fluctuations in its performance, which CPDNet should learn to detect.

To formulate the resulting loss measure used, we write the input features to CPDNet based on the innovation process produced by KalmanNet with parameters $\dnnParams$ applied to measurements $\myVec{y}_t^{(i)}$ as $\myVec{z}_t^{(i)}(\dnnParams)$. We can now treat CPDNet as a binary detection network, and train it based on the \ac{bce} loss evaluated over $\mySet{D}_{\rm v}$, namely
\begin{align}
    \mathscr{L}_{\mySet{D}_{\rm v}}^{\rm BCE}(\myVec{\phi}; \dnnParams) =
    &\frac{-1}{|\mySet{D}_{\rm v}|(T \!-\! \delta \!+ \!1)}
    \sum_{i=1}^{|\mathcal{D}_{\rm v}|}
    \sum_{t=\delta}^{T}
   p_t^{(i)}(\dnnParams) \log \hat{p}_{\myVec{\phi}}(\myVec{z}_t^{(i)}(\dnnParams)) \notag \\
    &\quad + (1 \!- \!p_t^{(i)}(\dnnParams)) \log(1 \!-\! \hat{p}_{\myVec{\phi}}(\myVec{z}_t^{(i)}(\dnnParams))) .
    \label{eqn:LossBCE}
\end{align}

The overall training procedure using \ac{sgd}-based learning is summarized as Algorithm~\ref{alg:Training}.

\begin{algorithm}
\caption{CPDNet Training using \ac{sgd}}
\label{alg:Training}
\SetKwInOut{Initialization}{Init}
\Initialization{Dataset $\mySet{D} = \mySet{D}_{\rm s}\cup \mySet{D}_{\rm v}$; Hyperparameters $\delta, \gamma$ \\ Step-size $\tilde{\eta}$, batches $L$, epochs $e_{\max}$; Initial  $\myVec{\phi}$}

\tcp{Stationary Reference Pre-Training}
Initialize reference KalmanNet $\dnnParams$\;
\For{${\rm epoch}=0,1,\ldots, e_{\max}$}
{
  Randomly divide  $\mathcal{D}_{\rm s}$ into $L$ batches $\{\mathcal{D}_{\rm s}^l\}_{l=1}^L$\;
      \For{$l = 1, \ldots, L$}{
        Compute KalmanNet \ac{mse} loss \eqref{eqn:LossMSE} on $\mathcal{D}_{\rm s}^l$\;
        Update $\dnnParams\leftarrow \dnnParams -\tilde{\eta} \nabla_{\dnnParams}\mathscr{L}_{\mySet{D}_{\rm s}^l}^{\rm MSE}(\dnnParams)$ \;
      }
}

\tcp{Nonstationary  Training}
\For{${\rm epoch}=0,1,\ldots, e_{\max}$}
{
  Randomly divide  $\mathcal{D}_{\rm v}$ into $L$ batches $\{\mathcal{D}_{\rm v}^l\}_{l=1}^L$\;
      \For{$l = 1, \ldots, L$}{
        Compute CPDNet \ac{bce} loss \eqref{eqn:LossBCE} on $\mathcal{D}_{\rm v}^l$\;
        Update $\myVec{\phi}\leftarrow \myVec{\phi} -\tilde{\eta} \nabla_{\myVec{\phi}}\mathscr{L}_{\mySet{D}_{\rm v}^l}^{\rm BCE}(\myVec{\phi};\dnnParams)$ \;
      }
}
\KwOut {Trained CPDNet $\myVec{\phi}$.}

\end{algorithm}

\section{\ac{name}}
\label{sec:Continual}
The neural \ac{cpd}  algorithm detailed in the previous section enables monitoring a trained KalmanNet. It provides a soft estimate which is indicative of both when KalmanNet starts producing errors, as well as of how large these errors are. In this section we show how the change awareness induced by CPDNet enables KalmanNet to be self-adaptive. We next present the resulting \ac{name} framework and its complexity, after which we provide a discussion.

 \subsection{Self-Adaptive KalmanNet}
 \label{subsec:continual}
 \subsubsection{Change Awareness}
In our self-adaptive framework, CPDNet is integrated into the overall KalmanNet processing chain, rather than act simply as an external monitoring mechanism. Specifically, on each incoming observation $\mathbf{y}_t$, KalmanNet computes the innovation process $\Delta \mathbf{y}_t$, which allows CPDNet to compute the features $\myVec{z}_t$ and produce its soft estimate $\hat{p}_{\myVec{\phi}}(\myVec{z}_t)$. This soft estimate is used to both {\em detect} whether adaptation should be activated and {\em modulate} the magnitude of the online update.

We formulate this mathematically by setting a time-varying learning-rate, denote $\eta_t$, which governs the adaptation of KalmanNet's parameters at time instance $t$. This learning rate is set as
\begin{equation}
    \label{eqn:TVLearningRate}
    \eta_t = \epsilon \cdot \max(\hat{p}_{\myVec{\phi}}(\myVec{z}_t) - {\rm thresh}, 0),
\end{equation}
where $\epsilon>0$ is a hyperparameter representing an upper bound on the maximal learning rate, and ${\rm thresh}$ is a threshold hyperparameter. When $\hat{p}_{\myVec{\phi}}(\myVec{z}_t) \leq {\rm thresh}$, the instantaneous learning rate is set to zero, which implies that no adaptation is carried out. In this work, an update is regarded as unnecessary in the operational sense when the CPDNet reliability-degradation score does not exceed this threshold. The ReLU--$\tanh$ output layer and the threshold in \eqref{eqn:TVLearningRate} therefore play different roles: the former constrains and biases the CPDNet score to the range $[0,1]$, whereas the latter is an explicit decision rule that converts the score into a zero or positive online learning rate. The resulting learning rate should thus be interpreted as a CPDNet-gated heuristic adaptation magnitude, rather than a theoretically guaranteed optimal update size.

\subsubsection{Online Learning}
We proceed to formulate the loss function that guides the instantaneous adaptation using the change-aware learning rate of \eqref{eqn:TVLearningRate}. Having identified the innovation process $\Delta \mathbf{y}_t$ (and particularly its magnitude) as a {ground-truth-free} indicator on the usefulness of KalmanNet to the underlying dynamics, we construct an unsupervised empirical loss based on this quantity. The resulting loss evaluated for KalmanNet with parameters $\dnnParams$ at time instance $t$ is given by  the $\ell_2$ regularized form
\begin{equation}
    \mathscr{L}^{\rm unsup}_{t}(\dnnParams) = \|\Delta \mathbf{y}_t(\dnnParams)\|^2 + \rho \|\dnnParams\|^2,
    \label{eqn:LossUnsup}
\end{equation}
where $\Delta \mathbf{y}_t(\dnnParams)$ represents the innovation process computed by KalmanNet with parameters $\dnnParams$ at time $t$, and $\rho$ is a regularization coefficient.
The loss employs the raw innovation magnitude rather than a normalized
form, as the true noise covariances
are unknown and time-varying in the online setting, rendering the
innovation covariance intractable. KalmanNet's offline pre-training
implicitly encodes scale-awareness, and the adaptive learning rate
$\eta_t$ in \eqref{eqn:TVLearningRate} absorbs residual magnitude discrepancies.

In formulating \eqref{eqn:LossUnsup}, we draw inspiration from the previous work \cite{revach2022unsupervised}, which also proposed an unsupervised loss measure based on the innovation process. It is noted, however, that \cite{revach2022unsupervised} considered {\em offline} learning from a complete unsupervised dataset using multiple iterations of \ac{sgd}-based optimization. In contrast, we consider online adaptation of a supervised pretrained KalmanNet, where the trainable parameters are updated from streaming observations using a single gradient step based on \eqref{eqn:LossUnsup} and \eqref{eqn:TVLearningRate}, i.e., the update rule at time $t$ is given by
\begin{equation}
    \dnnParams \leftarrow \dnnParams  - \eta_t \nabla_{\dnnParams}\mathscr{L}^{\rm unsup}_{t}(\dnnParams).
    \label{eqn:CASAGradStep}
\end{equation}
Computing the gradient in \eqref{eqn:CASAGradStep} employs \ac{bptt}. To maintain low complexity and avoid complexity growth with $t$, we employ the truncated version with $\tau_{\max}$ time steps \cite{sutskever2013training}.

The resulting online update is not intended to train KalmanNet from scratch, but rather to continuously adapt it to changes in the underlying dynamics. Its effect is constrained by three practical safeguards: the update is triggered only when CPDNet indicates a sufficiently large reliability degradation, the learning rate is bounded by the CPDNet score, and the unsupervised loss includes regularization to prevent excessive parameter drift. 


\subsubsection{Algorithm Summary}
\label{sssec:Algorithm Summary}

\begin{figure}
    \centering
    \includegraphics[width=\linewidth]{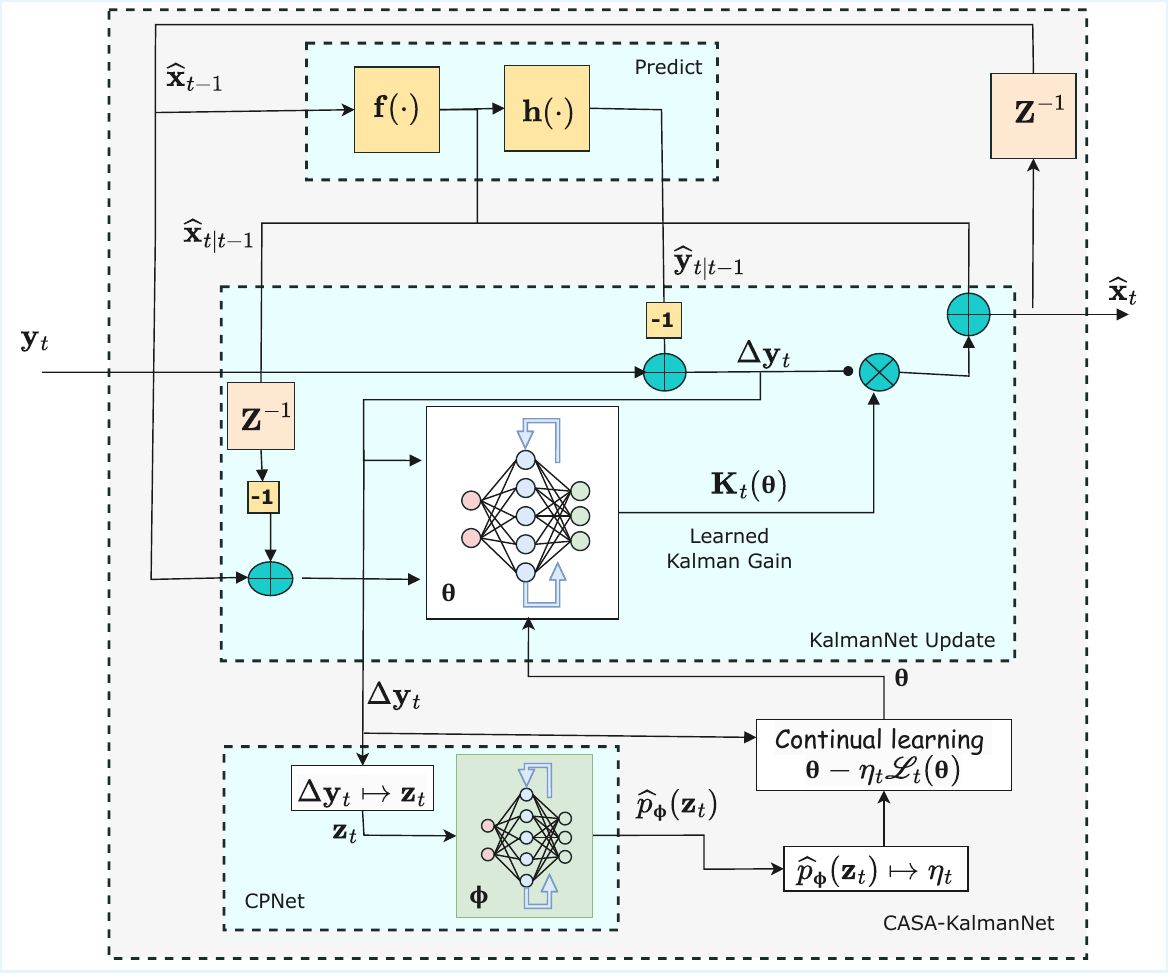}
    \caption{\ac{name} illustration.The framework combines KalmanNet with CPDNet-based online adaptation, where CPDNet extracts statistical features from the innovation to adjust the learning rate for updating KalmanNet parameters. $\mathbf{Z}^{-1}$ denotes the unit delay operator.}
    \label{fig:CASAKlamanNet}
\end{figure}

The overall \ac{name}, which extends KalmanNet to be self-adaptive by integrating CPDNet, is summarized as Algorithm~\ref{alg:cpdnet_guided_adaptation}, and illustrated in Fig.~\ref{fig:CASAKlamanNet}.

\begin{algorithm}
\caption{\ac{name} at time $t$}
\label{alg:cpdnet_guided_adaptation}
\SetKwInOut{Initialization}{Init}
\Initialization{Hyperparameters $\delta, \gamma, \alpha, \rho, \epsilon, \tau_{\max}$; \\
    \ac{ss} model functions $f(\cdot)$, $h(\cdot)$;\\
    CPDNet $\myVec{\phi}$, KalmanNet $\dnnParams$
}
\SetKwInOut{Input}{Input}
\Input{Observation $\mathbf{y}_t$; previous estimate $\hat{\mathbf{x}}_{t-1}$
}

    \tcp{KalmanNet predict}
    Predict prior state $\hat{\mathbf{x}}_{t|t-1} = f(\hat{\mathbf{x}}_{t-1})$\;
    Compute innovation process $\Delta{\mathbf{y}}_{t} = \mathbf{y}_t - h(\hat{\mathbf{x}}_{t|t-1})$\;

    \tcp{CPDNet}
    Construct CPDNet features $\myVec{z}_t$ from  $\Delta{\mathbf{y}}_{t}$ via \eqref{eqn:FeatureExt}\;
    Compute learning rate  $\eta_t$ from $\hat{p}_{\myVec{\phi}}(\myVec{z}_t)$ via \eqref{eqn:TVLearningRate}\;

    \tcp{Unsupervised update of KalmanNet}

    \If{$\eta_t > 0$}
    {
       Compute unsupervised loss $\mathscr{L}^{\rm unsup}_{t}(\dnnParams)$ via \eqref{eqn:LossUnsup}\;
        Adapt KalmanNet $\dnnParams \leftarrow \dnnParams  - \eta_t \nabla_{\dnnParams}\mathscr{L}^{\rm unsup}_{t}(\dnnParams)$\;
    }

    \tcp{KalmanNet update}
    Compute posterior: $\hat{\mathbf{x}}_{t} =\hat{\mathbf{x}}_{t|t-1} + \hat{\mathbf{K}}_t(\dnnParams) \Delta \mathbf{y}_t$\;

\KwRet{$\hat{\mathbf{x}}_{t}$}
\end{algorithm}

\subsection{Complexity Analysis}
\label{subsec:complexity_analysis}

We analyze the computational complexity of the proposed Algorithm~\ref{alg:cpdnet_guided_adaptation}. Specifically, the operation of \ac{name}
includes three main compute stages at each time instance $t$: $(i)$ applying CPDNet; $(ii)$ single-step training of KalmanNet (if needed); and $(iii)$ applying KalmanNet. By letting $\mySet{C}_{\rm inf}^{\rm CPD}$, $\mySet{C}_{\rm upd}^{\rm KNet}$, and $\mySet{C}_{\rm inf}^{\rm KNet}$ respectively denote the computational burden of $(i)-(iii)$, the overall per-time step complexity is given by
\begin{equation}
    \mySet{C}_{\rm CASA} = \mySet{C}_{\rm inf}^{\rm KNet}+ \mySet{C}_{\rm inf}^{\rm CPD} + \Pr(\eta_t > 0)\cdot \mySet{C}_{\rm upd}^{\rm KNet}
    \label{eqn:Complexity}
\end{equation}

To obtain a meaningful and tractable description of the excessive latency induced by \ac{name} compared to standard KalmanNet, i.e., $\Delta \mySet{C} \triangleq \mySet{C}_{\rm CASA} - \mySet{C}_{\rm inf}^{\rm KNet}$, we note that the complexity of truncated \ac{bptt} scales linearly with the number of trainable parameters times the (truncated) sequence length~\cite{gruslys2016memory}, and thus we write $\mySet{C}_{\rm upd}^{\rm KNet} = \gamma_{\rm upd} \cdot \tau_{\max} \cdot {\rm len}(\myVec{\theta})$ for some $\gamma_{\rm upd} > 0$, where ${\rm len}( \cdot)$ applied to a vector returns its length.
Using this introduced notation, the excessive complexity becomes
\begin{equation}
    \Delta \mySet{C} = \mySet{C}_{\rm inf}^{\rm CPD} +  \Pr(\eta_t > 0)\cdot \gamma_{\rm upd} \cdot \tau_{\max} \cdot {\rm len}(\myVec{\theta}).
    \label{eqn:ExcessiveComp}
\end{equation}

While the excessive latency in \eqref{eqn:ExcessiveComp} is given in parametric form, it allows to qualitatively assess the overhead induced by \ac{name}. Specifically, CPDNet is comprised of a single-layer \ac{rnn} applied to a sequence with limited length $\delta$, and thus the overhead associated with $\mySet{C}_{\rm inf}^{\rm CPD}$ is limited. The complexity of updating KalmanNet depends on the predefined $\tau_{\max}$ and on the probability of triggering an update, which can be controlled by the threshold hyperparameter ${\rm thresh}$.
Here, $\delta$ and $\tau_{\max}$ have different meanings. The former is the CPDNet detection window, while the latter is the truncated backpropagation horizon used to bound the computational cost of each online KalmanNet update. In practice, both should be chosen to be shorter than the expected duration of a stationary regime, so that CPDNet can detect local reliability degradation and the update remains localized to the current regime. A larger $\tau_{\max}$ allows the update to use longer temporal information but increases latency, whereas a smaller $\tau_{\max}$ reduces overhead at the cost of a more myopic correction. This results in \eqref{eqn:ExcessiveComp} often translating to limited and even negligible excessive latency while providing self-adaptivity.

 \subsection{Discussion}
 \label{subsec: Discussion}

The proposed \ac{name} framework is designed to address the key challenges \ref{itm:Approx}--\ref{itm:varying} associated with non-stationary dynamical environments, as formulated in Subsection~\ref{ssec:Problem}. By integrating a neural-based \ac{cpd} module (CPDNet) with an online adaptation learning strategy, \ac{name} preserves the capability of KalmanNet to handle approximate models~(\ref{itm:Approx}) and unknown noise distributions~(\ref{itm:Noise}), while further enabling timely and efficient adaptation~(\ref{itm:varying}). Unlike conventional pre-trained \ac{ai}-aided Kalman filters, the integration of CPDNet allows the system to actively monitor its own reliability and autonomously determine \emph{when} and \emph{to what extent} adaptation should occur, thereby effectively mitigating both under- and over-adaptation.

The current design of \ac{name} is tailored towards the KalmanNet \ac{ai}-aided framework, where the dominant source of variability arises from changes in the noise statistics. Moreover, our numerical experiments in Section~\ref{sec: Numerical Study} demonstrate that \ac{name} can also cope with some level of variations in the state-transition and observation functions. This capability leverages KalmanNet’s established robustness to moderate model mismatches. Nevertheless, extending the methodology to scenarios involving significant structural changes in the system dynamics would likely require an additional layer of system identification, since KalmanNet relies on accurate state-transition and observation models to compute the innovation term $\Delta \myVec{y}_t$, which is central to our unsupervised loss formulation. Furthermore, while \ac{name} is inherently suited for \ac{ai}-aided Kalman filters due to their interpretable internal structure, the proposed methodology could potentially inspire adaptations to other neural-based architectures. Achieving this, however, would require redesigning the feature-monitoring mechanism and rethinking the loss formulation to exploit the internal representations specific to those models.

Finally, this work opens several promising research directions. First, incorporating Bayesian deep learning into \ac{name} may provide principled uncertainty quantification~\cite{dahan2024uncertainty}, which can be leveraged for more reliable and calibrated self-adaptation. Second, dynamically adjusting the number of gradient steps and the observation window length based on the severity of the detected change could better balance adaptation responsiveness with computational overhead. These extensions are left for future investigation.


\section{Numerical Study}
\label{sec: Numerical Study}

 In this section, we numerically evaluate the effectiveness of our self-adaptive methodology,  assessing the ability of CPDNet to detect reliability degradation and how well \ac{name} adapts to abrupt and gradual changes in the underlying \ac{ss} model. To that aim, we first evaluate only CPDNet, focusing on linear Gaussian systems, where changes in the \ac{ss} model are manually injected at unknown time steps to simulate non-stationary environments. Then we proceed to evaluate the overall online adaptation framework in both linear settings as well as nonlinear dynamics, using the Lorenz attractor \ac{ss} model. We also evaluate our method on the real-world NCLT robot dataset~\cite{ncarlevaris-2015a}.
Unless otherwise specified, all filters are initialized with the same state estimate and covariance matrix for fair comparison. Specifically, we set the initial state estimate as $\hat{\mathbf{x}}_{0|0}=\mathbf{0}$ and use the same fixed positive definite matrix $\boldsymbol{\Sigma}_{0|0}$ as the initial error covariance for all filters that require covariance initialization. The simulated trajectories are also generated from a zero initial state, assuring that the reported performance differences are not caused by different initialization choices across methods. 
Throughout our experiments, we use KalmanNet based on Architecture 2 of \cite{revach2022kalmannet}, and our source code is available online at \url{https://github.com/ZhangWenyi01/CASA-KalmanNet}.

\subsection{Change Point Detection Evaluation}
\label{subsec:CPDNetSim}
\subsubsection{Setup}
To thoroughly evaluate the robustness and adaptive capability of  CPDNet under varying conditions, we  design experiments involving abrupt changes or gradual changes in different system parameters. Specifically,  we simulate a linear \ac{ss} models as in \eqref{eq:ss_linear} with Gaussian noises. The state transition matrix $\mathbf{F}$ and the observation matrix $\mathbf{H}$ are assumed to be known, while changes are introduced in the noise covariances $\mathbf{Q}$ and $\mathbf{R}$, and optionally in the system matrices themselves.
 The simulated \ac{ss} model is given by
\begin{equation}
\label{eqn:system_matrices_aligned}
\begin{aligned}
\mathbf{F} &=
\begin{bmatrix}
1 & \Delta t & \tfrac{1}{2} \Delta t^2 \\
0 & 1 & \Delta t \\
0 & 0 & 1
\end{bmatrix}, &
\mathbf{H} &=
\begin{bmatrix}
1 & 0 & 0
\end{bmatrix}, \\
\mathbf{Q}_1 &=
\begin{bmatrix}
\tfrac{1}{20} \Delta t^5 & \tfrac{1}{8} \Delta t^4 & \tfrac{1}{6} \Delta t^3 \\
\tfrac{1}{8} \Delta t^4 & \tfrac{1}{3} \Delta t^3 & \tfrac{1}{2} \Delta t^2 \\
\tfrac{1}{6} \Delta t^3 & \tfrac{1}{2} \Delta t^2 & \Delta t
\end{bmatrix}, &
\mathbf{R}_1 &=
\begin{bmatrix}
1
\end{bmatrix}.
\end{aligned}
\end{equation}
{This \ac{ss} model represents one-dimensional target movement based on the
constant acceleration model~\cite{bar2004estimation}.
The state elements represent position, velocity, and acceleration.}

In \eqref{eqn:system_matrices_aligned}, the term $\Delta t$ represents the sampling interval, which is set to $0.01$s for this experiment. This \ac{ss} model represents one-dimensional target movement based on the constant acceleration model~\cite{bar2004estimation}, with the state elements representing position, velocity, and acceleration.

We simulate two forms of changes:
\begin{itemize}
\item {\em Abrupt changes}, where in a sequence of length equals  $T$, we insert a random change point $t_{\rm change} \sim \mathcal{U}[1,T]$ within each generated sequence. The change point separates the pre-change stationary regime $t\leq t_{\rm change}$ from the post-change stationary regime $t>t_{\rm change}$, where the varying parameter scales by a coefficient $\lambda$ across the boundary.
For example, abrupt changes in $\mathbf{Q}$ are set as
\begin{equation}
\label{eqn:abrupt}
    \mathbf{Q}_t = \begin{cases}
        \mathbf{Q}_1 & t \leq t_{\rm change} \\
        \lambda \cdot \mathbf{Q}_1 & t > t_{\rm change}
    \end{cases}.
\end{equation}

    \item {\em Gradual changes}, where the varying parameter scales on each time instance. For example, gradual changes in $\mathbf{Q}$  are
\begin{equation}
\label{eqn:gradual}
 \mathbf{Q}_t= \begin{cases}
            \mathbf{Q}_1 & t \leq t_{\rm change} \\
            \lambda^{t-t_{\text{change}}} \cdot \mathbf{Q}_1 & t > t_{\rm change}
    \end{cases}.
\end{equation}
\end{itemize}
We note that the entries of $\mathbf{Q}_1$ are numerically small mainly due to the discretization of the constant-acceleration model with a small sampling interval $\Delta t=0.01$, where the covariance terms scale with powers of $\Delta t$. Therefore, the absolute entries of $\mathbf{Q}_1$ should not be interpreted as a direct scalar comparison with the observation covariance $\mathbf{R}=[1]$, since the two matrices act on different spaces and have different physical units. Although the nominal process noise is smaller than the observation noise in this discretized setting, the post-change regimes considered in Table~\ref{tab:cpdnet_results} include large process-noise scalings, which produce a clear mismatch relative to the stationary training regime.


\subsubsection{Results}
We evaluate CPDNet by comparing its predictions with the error signals obtained from KalmanNet. Note that these errors are not actual probabilities, but rather probabilistic representations of the state estimation error produced by KalmanNet. For clarity, we refer to them as KalmanNet errors throughout the paper. In our experiments, the process noise covariance ($\mathbf{Q}$) is adjusted by setting $\lambda \in \{100, 300, 500\}$, representing dominant abrupt changes, as well as by gradual changes. Similarly, the measurement noise covariance ($\mathbf{R}$) is varied with less dominant variations $\lambda \in \{1.2, 1.3, 1.5\}$ and also with gradual transitions. These settings allow us to assess how sensitively CPDNet detects changes when the noise characteristics vary. For the state transition matrix ($\mathbf{F}$) and observation matrix ($\mathbf{H}$), we introduce minor but impactful deviations by scaling them with factors $\lambda \in \{1.01, 0.95\}$ for $\mathbf{F}$ and $\lambda \in \{1.01, 0.95\}$ for $\mathbf{H}$, respectively, modeling practical cases with minor unknown shifts in the model.

\begin{table}
\caption{Discrepancy between CPDNet predictions
and KalmanNet errors}
\label{tab:cpdnet_results}
\centering
\begin{tabular}{|c|c|c|}
\hline
\textbf{Time-varying parameter} & \textbf{$\lambda$} & \textbf{MSE (dB)} \\
\hline
\multirow{4}{*}{$\mathbf{Q}$}
  & 100       & -26.255 \\
  & 300       & -22.203 \\
  & 500       & -18.175 \\
  & Gradually & -20.661 \\
\hline
\multirow{4}{*}{$\mathbf{R}$}
  & 1.2       & -23.753 \\
  & 1.3       & -23.055 \\
  & 1.5       & -20.668 \\
  & Gradually & -19.712 \\
\hline
\multirow{2}{*}{$\mathbf{F}$}
  & 1.01      & -23.333 \\
  & 0.95      & -17.858 \\
\hline
\multirow{2}{*}{$\mathbf{H}$}
  & 1.01       & -25.731 \\
  & 0.95      & -17.858 \\
\hline
\end{tabular}
\end{table}

\begin{figure}
\centering
\includegraphics[width=1\linewidth]{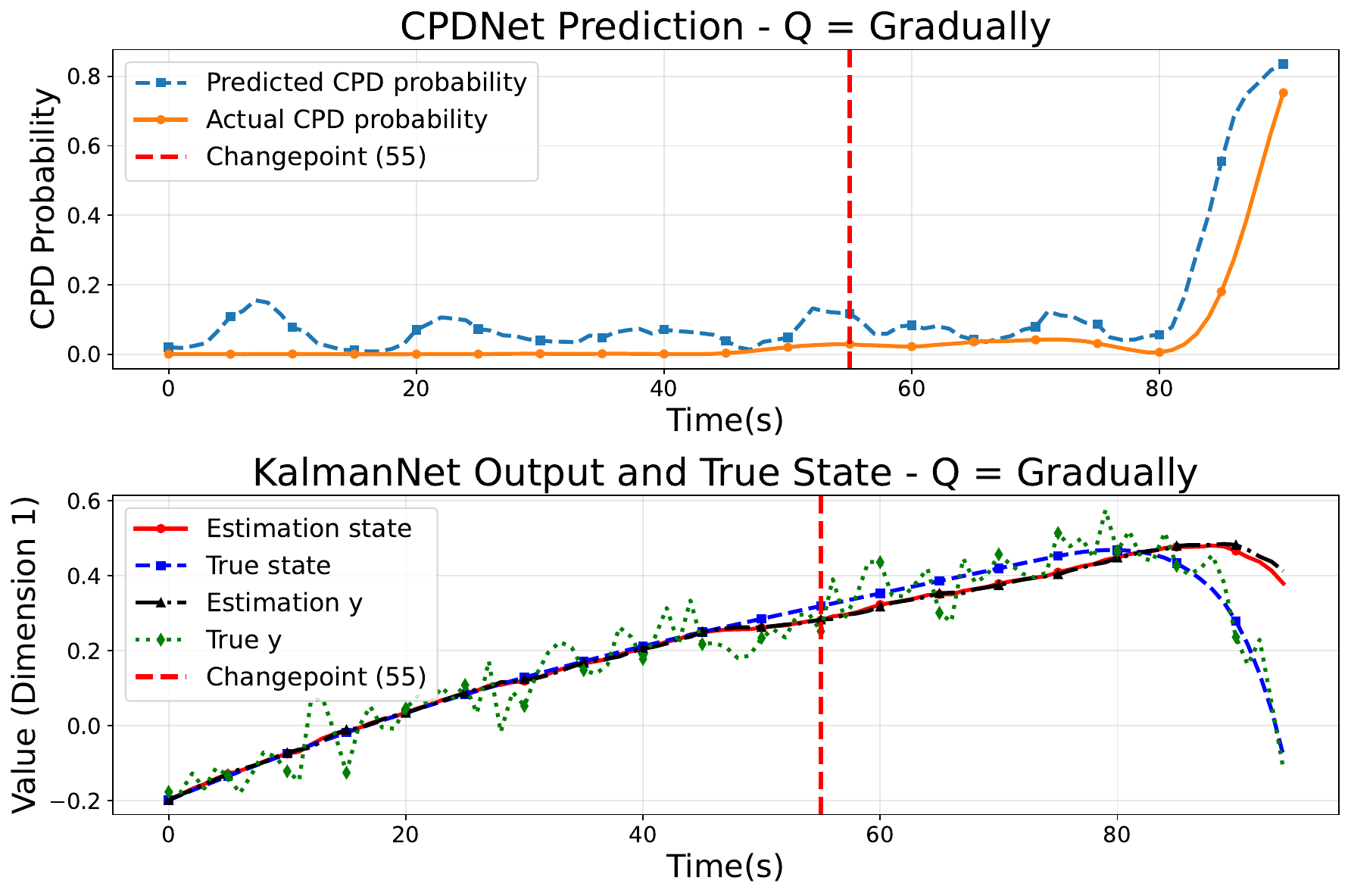}
\caption{Performance of CPDNet under gradual changes in process noise covariance $\mathbf{Q}$. The top panel shows the predicted and actual \ac{cpd} probabilities, together with the ground-truth change point. The bottom panel compares the KalmanNet estimation and the true state/output for the first component of the state vector, plotted in arbitrary units (a.u.) to indicate that no fixed physical units are associated with the values. }
\label{fig:change_Q_grad}
\end{figure}

\begin{figure}
\centering
\includegraphics[width=1\linewidth]{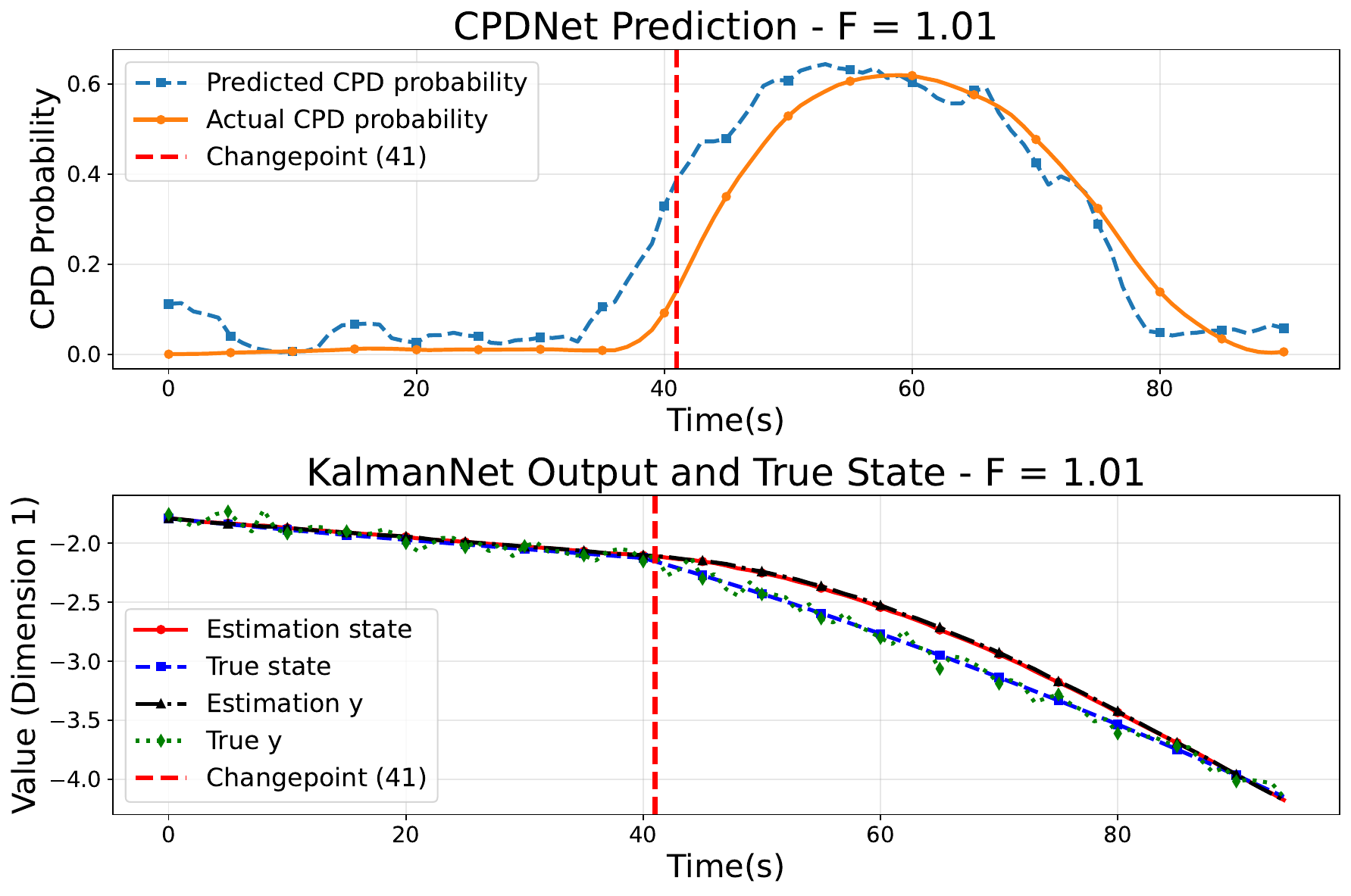}
\caption{Performance of CPDNet under $\lambda=1.01$ change in state transition matrix $\mathbf{F}$, with the same plotting format as in Fig. \ref{fig:change_Q_grad}.}
\label{fig:change_F_1.01}
\end{figure}

The overall \ac{mse} results  listed in Table~\ref{tab:cpdnet_results} illustrate CPDNet's effectiveness and stability, explicitly demonstrating its ability to maintain high accuracy and quickly adapt in the presence of abrupt parameter changes. Specifically, the results reveal how accurately CPDNet identifies the introduced change points and predicts the estimation performance, with only minor errors in the predicted error.
Moreover, Figs.~\ref{fig:change_Q_grad} and \ref{fig:change_F_1.01}
depict representative scenarios of \ac{cpd} at specific trajectories. These figures correspond to selected parameter changes from Table~\ref{tab:cpdnet_results}. In each figure, the upper sub-figure shows the predicted error produced by CPDNet compared to the actual error (mapped into probabilistic value via \eqref{eqn:Label}), while the lower sub-figure depicts the temporal behavior of the state (estimated by KalmanNet and true state) as well as the observation $\mathbf{y}_t$ (both the observation predicted by KalmanNet and the true value). These figures systematically shows that CPDNet spikes in the time step where the error of KalmanNet becomes dominant and necessitates adaptation. 

\subsection{Linear Gaussian State-Space Model}
\label{subsec:linear state space model}
\subsubsection{Setup}
Having numerically asserted the ability of CPDNet to reliably predict the error behavior of KalmanNet, we proceed to evaluate its integration into the overall online adaptation framework of \ac{name}. We again consider the linear \ac{ss} model as in Subsection~\ref{subsec:CPDNetSim}.  Here, the main evaluation metric is the \ac{mse} between the estimated states and the ground truth:
\begin{equation}
    \text{MSE (dB)} = 10 \log_{10} \left( \frac{1}{T} \sum_{t=1}^T \| \hat{\mathbf{x}}_t - \mathbf{x}_t \|^2 \right).
    \label{Eqn:TAMSE}
\end{equation}
The time-averaged \ac{mse} \eqref{Eqn:TAMSE} is used in following experiments to compactly compare the overall estimation accuracy across multiple change scenarios and competing methods. 
Time-varying behavior is illustrated through the CPDNet output and the corresponding KalmanNet error in Fig.~\ref{fig:change_F_1.01}, while averaged \ac{mse} is used for concise cross-scenario comparison.
All experiments are conducted on sequences of length $T=100$, with change points randomly inserted.  This controlled design enables a fair comparison across methods and highlights the benefits of online adaptation for online state estimation.

We compare the following algorithms:
\begin{itemize}
    \item \textbf{{KalmanNet (Supervised)}}: This model is pre-trained with full supervision on trajectories generated from a static \ac{ss} model, which does not contain change points. It is deployed in a fixed manner without any subsequent online training or adaptation.
    \item \textbf{{KalmanNet (Unsupervised)}}: This variant is also pre-trained on trajectories from a static system without change points. However, during the inference phase, it continuously adapts its parameters online on each time instance by minimizing the one-step prediction error, without requiring access to the ground truth state.

    \item \textbf{\ac{name} (Ours)}: Integrates CPDNet to dynamically adjust the learning rate of the unsupervised KalmanNet, guided by detected change points.
    \item \textbf{\ac{kf}}: The standard \ac{kf} operating based on pre-defined \ac{ss} model parameters, which can be accurate (full information) or mismatched (partial information).
    \item \textbf{Expectation-Maximization Kalman Filter (EM-KF)}: The EM-KF~\cite{Shumway1982} is employed to address inaccuracies in the state noise covariance matrix $\mathbf{Q}$ and is only included in the experiments involving changes in $\mathbf{Q}$.
\end{itemize}

All methods provide causal estimates of $\mathbf{x}_t$ at time $t$, except EM-KF, which is a smoothing-based adaptive reference. Specifically, EM-KF is given the change time $t_{\rm change}$ and estimates the post-change trajectory only after observing the segment $[t_{\rm change},T]$. Thus, EM-KF is not a strictly causal competitor, but a strong model-based reference for assessing the gap between neural online adaptation and explicit parameter re-estimation.

\subsubsection{Full information results}
\label{sssec:full information}

We first consider the setting where the parameters of the \ac{ss} model are fully known. We simulate abrupt changes in the system, specifically in the process noise covariance $\mathbf{Q}$, measurement noise covariance $\mathbf{R}$, state transition matrix $\mathbf{F}$, and observation matrix $\mathbf{H}$, as in Subsection~\ref{subsec:CPDNetSim}.
The change time is assumed to be known to both the \ac{kf} and the EM-KF. However, while the \ac{kf} is provided with the updated parameter values after the change, KalmanNet and the EM-KF do not have access to these new values, with \ac{name} autonomously learning to adapt to such variations, while EM-KF  instead estimates them online via the EM procedure.

The MSE across all dimensions of the state vector, including position, velocity, and acceleration, are shown in Table~\ref{tab:all dim}.
Here, $1/r^2$ denotes the inverse observation noise variance in dB, where the observation noise covariance is parameterized as $\mathbf{R}=r^2\mathbf{I}$. Therefore, $1/r^2=0$~dB corresponds to $r^2=1$; in the one-dimensional observation setting of Subsection~\ref{subsec:CPDNetSim}, this reduces to $\mathbf{R}=[1]$.
As expected, the \ac{kf} and EM-KF with full parameter knowledge achieve the best overall performance among all tested methods, with the EM-KF demonstrating superior results compared to the standard KF due to its additional smoothing capabilities. Meanwhile, our proposed \ac{name} significantly outperforms all other hybrid-driven methods and approaches the performance of these theoretically optimal full-information filters. Precisely, \ac{name} achieves an MSE of $12.54$~dB at 0 dB signal-
to-noise ratios(SNR), substantially outperforming KalmanNet ($13.37$~dB) and Unsupervised KalmanNet ($22.61$~dB), demonstrating the effectiveness of our approach even without access to full system parameters.

\begin{figure}
    \centering
    \includegraphics[width=0.8\linewidth]{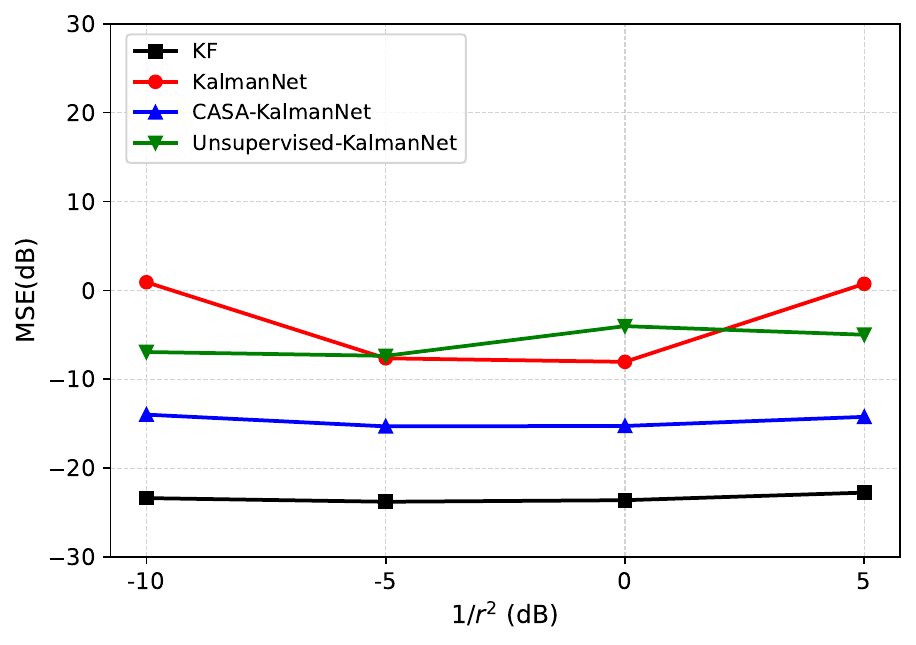}
    \caption{Linear model result on $\myMat{H}$ rotate after change}
    \label{fig:linear full H}
\end{figure}

\begin{figure}
    \centering
    \includegraphics[width=1\linewidth]{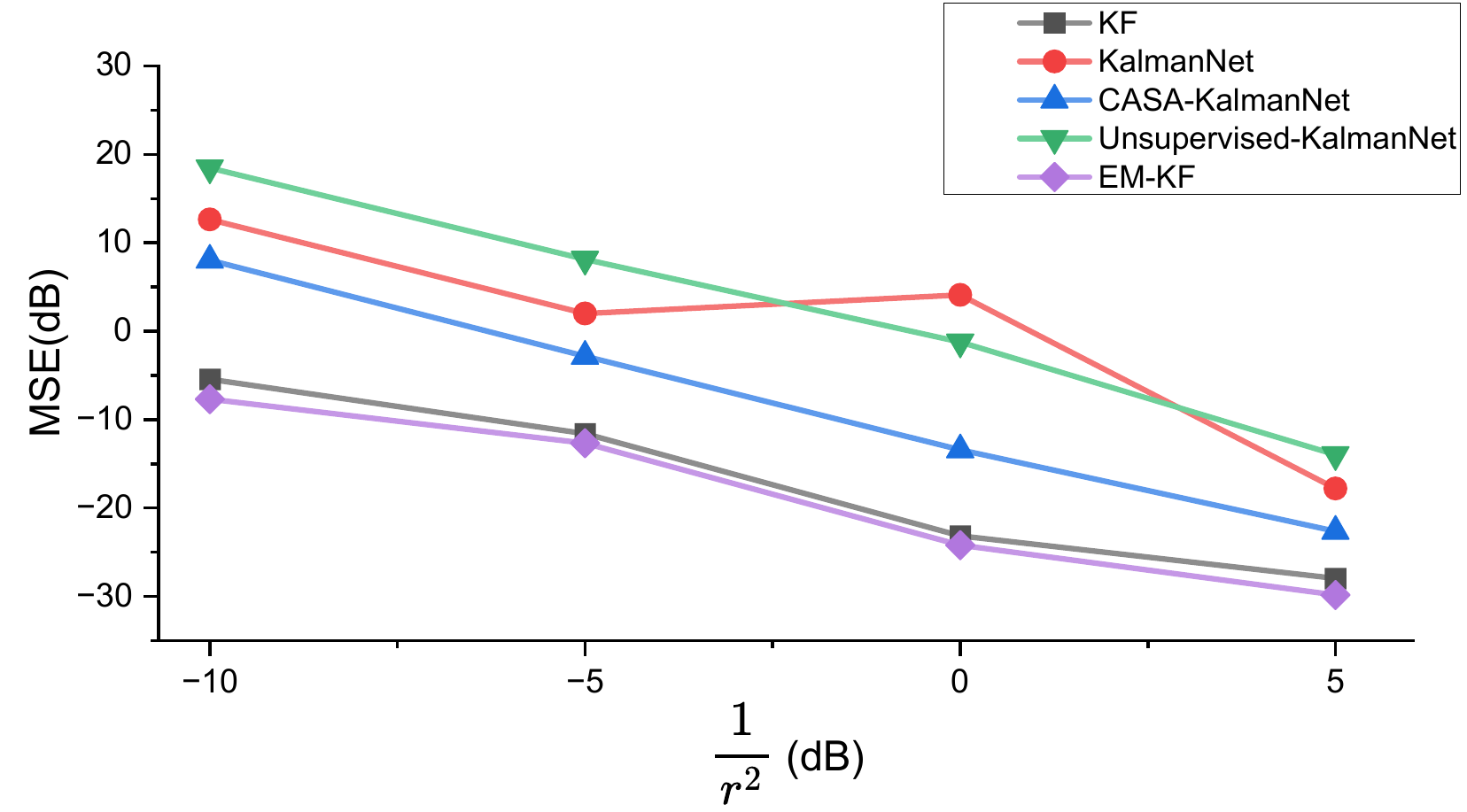}
    \caption{Linear model result on $\myMat{Q}$ change gradually after change}
    \label{fig:linear full Q}
\end{figure}

We proceed to evaluate additional forms of variations in the \ac{ss} model.
For ease of presentation, in the subsequent experiments, we focus exclusively on the position dimension (first state element) MSE.
The experimental results are presented in Table~\ref{tab:combined_metrics} and Figs.~\ref{fig:linear full H} and \ref{fig:linear full Q}. These figures report the \ac{mse} performance metrics across various conditions, including abrupt and gradual changes in $\mathbf{Q}$ and $\mathbf{R}$ matrices, as well as rotational transformations of the $\mathbf{F}$ and $\mathbf{H}$ matrices.
Specifically, for $\mathbf{F}$ and $\mathbf{H}$, we introduce a 2D rotation matrix $\mathbf{T}$ around a fixed axis, defined as
\[
\mathbf{T} =
\begin{bmatrix}
\cos\psi & -\sin\psi & 0 \\
\sin\psi & \ \cos\psi & 0 \\
0 & 0 & 1
\end{bmatrix},
\]
where $\psi$ denotes the rotation angle. The transformed matrices are then obtained via
$\mathbf{F}_{\mathrm{rot}} = \mathbf{T} \mathbf{F} \mathbf{T}^\top$ and
$\mathbf{H}_{\mathrm{rot}} = \mathbf{H} \mathbf{T}^\top$.
The performance is evaluated under different noise conditions, characterized by the inverse noise variance parameter $\frac{1}{r^2}$ ranging from $-10$~dB to $5$~dB.


The results
in Table~\ref{tab:combined_metrics} and Figs.~\ref{fig:linear full H} and \ref{fig:linear full Q}
demonstrate that given full domain knowledge, the \ac{kf} and EM-KF with full parameter knowledge achieve superior performance across all experimental scenarios.
Among the hybrid-driven approaches, \ac{name} consistently outperforms the competing methods across all tested conditions. The method demonstrates robust performance with relatively stable error metrics, particularly in scenarios involving parameter changes and matrix rotations. Under gradual changes in $\myMat{Q}$  scenario at 0~dB, \ac{name} achieves $-13.40$~dB compared to EM-KF's $-24.19$~dB and KF's $-23.13$~dB, retaining over $55\%$ of the performance gap between hybrid-driven and optimal methods.
%
These findings indicate that the proposed \ac{name}  achieves superior adaptability and robustness compared to existing hybrid-driven filtering approaches while maintaining competitive performance relative to classical methods that operate with full domain knowledge in the setting for which they are designed, of linear Gaussian dynamics. The consistent performance advantage over other hybrid-driven methods, combined with the ability to approach theoretically optimal full-information filters, demonstrates the effectiveness of our proposed approach.

\begin{table}
    \centering
    \caption{Algorithm MSE on all dimensions}
    \label{tab:all dim}
    \begin{tabular}{|l|c|c|c|c|}
        \hline
        \(1/r^2 \, [\text{dB}]\) & \multicolumn{1}{c|}{$-10$} & \multicolumn{1}{c|}{$-5$} & \multicolumn{1}{c|}{$0$} & \multicolumn{1}{c|}{$5$} \\
        \hline
        KalmanNet & 37.74 & 39.52 & 13.37 & 27.27 \\
        \hline
        Unsupervised & 61.64 & 27.33 & 22.61 & 24.10 \\
        \hline
        CASA-KalmanNet & 28.04 & 18.30 & 12.54 & 20.03 \\
        \hline
        KF & 24.48 & 15.01 & 11.35 & 18.98 \\
        \hline
        EM-KF & 23.09 & 14.37 & 10.78 & 19.53 \\
        \hline
    \end{tabular}
\end{table}

\begin{table*}
\centering
\caption{Time-averaged \ac{mse} across various non-stationarities}
\label{tab:combined_metrics}
\small 
\setlength{\tabcolsep}{6pt} 
\renewcommand{\arraystretch}{1.2} 

\begin{tabular}{|l|r|r|r|r|r|r|r|r|}
\hline
\multirow{2}{*}{\textbf{Method}} &
\multicolumn{4}{c|}{\cellcolor{gray!20}\textbf{Q Change Abruptly}} &
\multicolumn{4}{c|}{\cellcolor{gray!20}\textbf{Q Change Gradually}} \\
\cline{2-9}
& \textbf{-10} & \textbf{-5} & \textbf{0} & \textbf{5} &
  \textbf{-10} & \textbf{-5} & \textbf{0} & \textbf{5} \\
\hline

KF           & $-0.86$  & $-14.62$ & $-17.87$ & $-6.33$  & $-5.44$  & $-11.58$ & $-23.13$ & $-27.95$ \\
\hline
KalmanNet    & $16.68$  & $18.23$  & $-6.40$  & $8.08$   & $12.63$  & $2.00$   & $4.11$   & $-17.79$ \\
\hline
\rowcolor{blue!10}
CASA KalmanNet         & $\mathbf{6.57}$   & $\mathbf{-2.72}$  & $\mathbf{-11.44}$ & $\mathbf{-2.84}$  & $\mathbf{8.03}$   & $\mathbf{-2.82}$  & $\mathbf{-13.40}$ & $\mathbf{-22.62}$ \\
\hline
Unsupervised & $39.12$  & $5.74$   & $0.88$   & $7.00$   & $18.47$  & $8.12$   & $-1.23$  & $-13.96$ \\
\hline
EM-KF        & $-3.79$  & $-16.22$ & $-19.15$ & $-5.76$  & $-7.67$  & $-12.66$ & $-24.19$ & $-29.82$ \\
\hline\hline

\multirow{2}{*}{\textbf{Method}} &
\multicolumn{4}{c|}{\cellcolor{gray!20}\textbf{R Change Abruptly}} &
\multicolumn{4}{c|}{\cellcolor{gray!20}\textbf{R Change Gradually}} \\
\cline{2-9}
& \textbf{-10} & \textbf{-5} & \textbf{0} & \textbf{5} &
  \textbf{-10} & \textbf{-5} & \textbf{0} & \textbf{5} \\
\hline

KF           & $-7.16$  & $-14.11$ & $-22.19$ & $-28.23$ & $-6.63$  & $-11.86$ & $-18.48$ & $-28.56$ \\
\hline
KalmanNet    & $15.56$  & $15.76$  & $24.08$  & $-14.06$ & $22.12$  & $60.75$  & $-2.20$  & $-8.95$ \\
\hline
\rowcolor{blue!10}
CASA KalmanNet         & $\mathbf{7.42}$   & $\mathbf{-2.10}$  & $\mathbf{-13.04}$ & $\mathbf{-22.96}$ & $\mathbf{15.47}$  & $\mathbf{2.36}$   & $\mathbf{-12.72}$ & $\mathbf{-20.98}$ \\
\hline
Unsupervised & $15.76$  & $4.71$   & $2.39$   & $31.72$  & $26.14$  & $26.00$  & $10.12$  & $-12.77$ \\
\hline\hline

\multirow{2}{*}{\textbf{Method}} &
\multicolumn{4}{c|}{\cellcolor{gray!20}\textbf{F Rotated}} &
\multicolumn{4}{c|}{\cellcolor{gray!20}\textbf{H Rotated}} \\
\cline{2-9}
& \textbf{-10} & \textbf{-5} & \textbf{0} & \textbf{5} &
  \textbf{-10} & \textbf{-5} & \textbf{0} & \textbf{5} \\
\hline

KF           & $-18.02$ & $-19.96$ & $-26.52$ & $-30.99$ & $-23.37$& $-23.78$& $-23.61$& $-22.75$\\
\hline
KalmanNet    & $17.44$  & $8.22$   & $-6.84$  & $-10.15$ & $0.9178$& $-7.64$& $-8.04$& $0.73$\\
\hline
\rowcolor{blue!10}
CASA KalmanNet         & $\mathbf{6.94}$   & $\mathbf{-2.32}$  & $\mathbf{-11.61}$ & $\mathbf{-21.05}$ & $\mathbf{-13.97}$& $\mathbf{-15.3}$& $\mathbf{-15.25}$& $\mathbf{-14.22}$\\
\hline
Unsupervised & $36.60$  & $3.93$   & $-3.81$  & $-15.04$ & $-6.93$& $-7.36$& $-4.02$& $-4.99$\\
\hline
\end{tabular}

\medskip
\footnotesize
\end{table*}


\subsubsection{Partial information results}
\label{sssec:full information}

\begin{figure}
    \centering
    \includegraphics[width=0.85\linewidth]{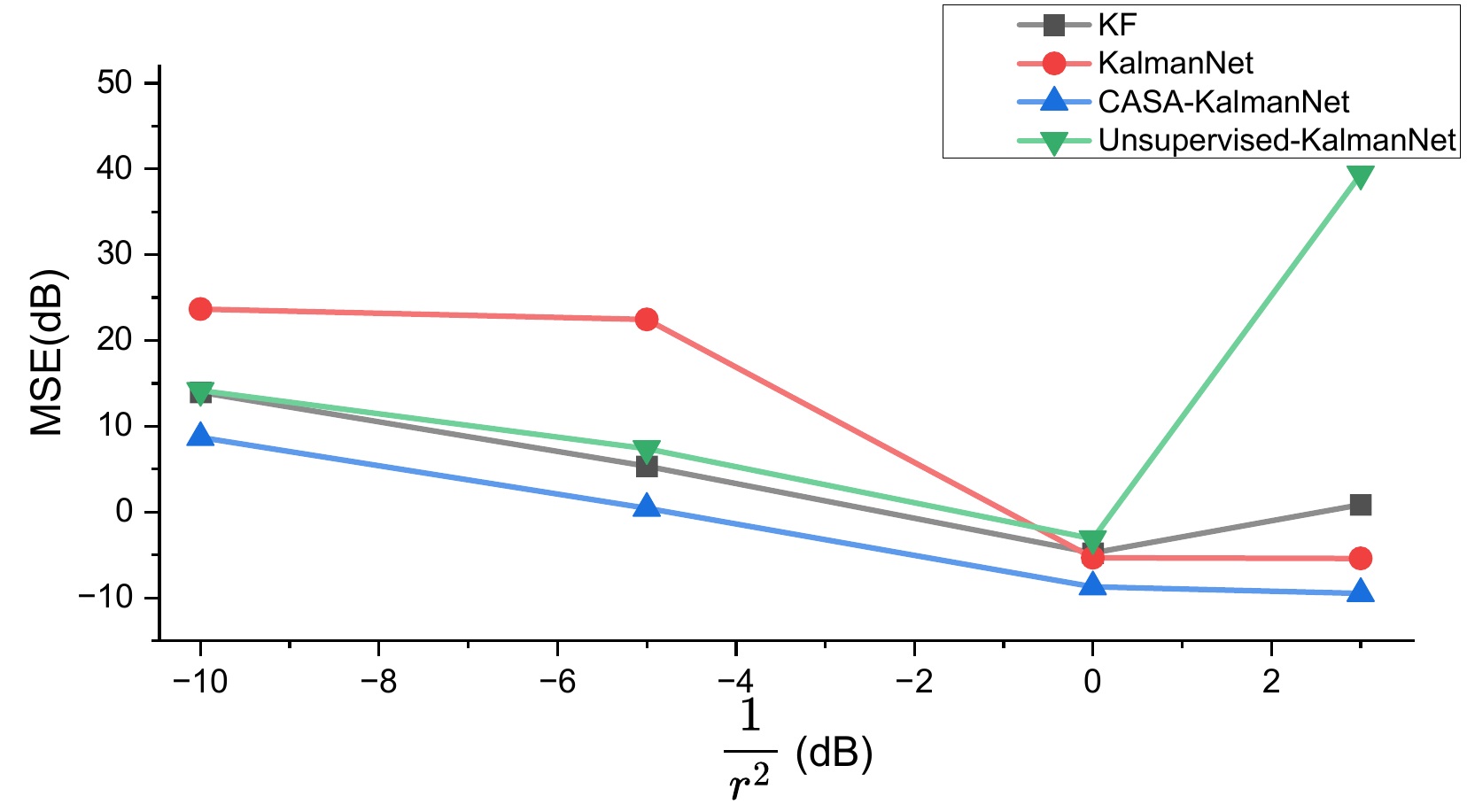}
    \caption{Position MSE performance under partial information scenario (Case~1), where the change point and post-change parameters are unknown.}
    \label{fig:partial_info_case1}
\end{figure}

We proceed  to evaluate \ac{name} and the baseline methods under partial information settings, where the generative parameters are only partially known to the filtering algorithms. Two experimental configurations are considered, both assuming that the process noise covariance matrix $\mathbf{Q}$ undergoes an abrupt change at the change point.
In the first experiment ({\em Case~1}), all models are assumed to have exact knowledge of the initial parameters $\mathbf{Q}$, $\mathbf{R}$, $\mathbf{F}$, and $\mathbf{H}$, but they are unaware of the change point location and the post-change parameter values. This setting examines the ability of each method to adapt to unexpected and significant process noise changes without prior knowledge of when or how these changes occur. As shown in Fig.~\ref{fig:partial_info_case1}, \ac{name} consistently outperforms the other hybrid-driven approaches across all SNR levels. For instance, at $0$~dB,  \ac{name} achieves an MSE of $-8.71$~dB, surpassing Supervised KalmanNet ($-5.33$~dB) and the Unsupervised KalmanNet ($-3.12$~dB), while remaining close to the partial information KF ($-4.76$~dB). At high-noise conditions ($-10$~dB),  \ac{name} reaches $8.71$~dB compared to $13.96$~dB for KF and $23.66$~dB for KalmanNet, demonstrating strong robustness in severe noise environments.

\begin{figure}
    \centering
    \includegraphics[width=0.85\linewidth]{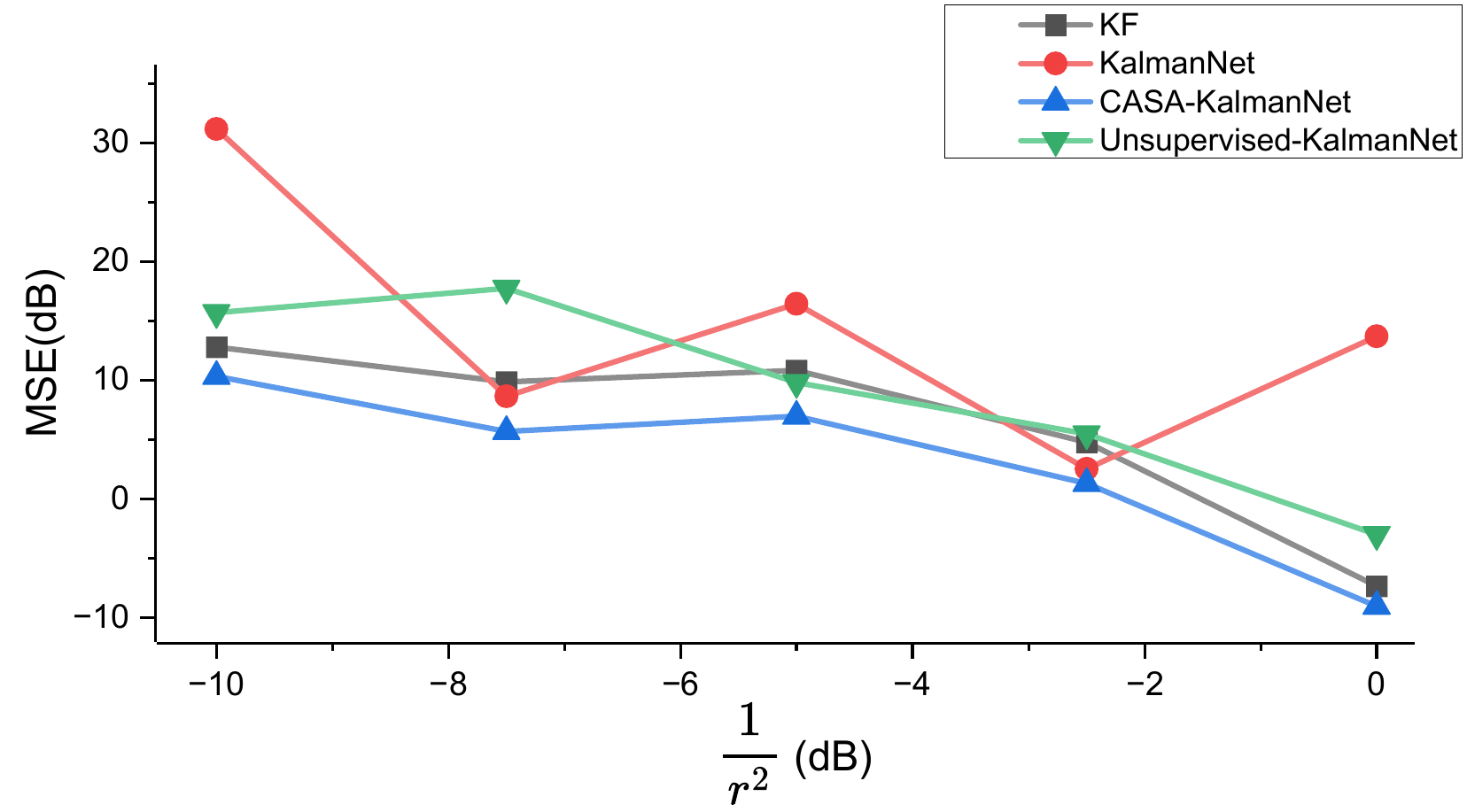}
    \caption{Position MSE performance under partial information scenario (Case~2), with both observation model mismatch and unknown change point.}
    \label{fig:partial_info_case2}
\end{figure}

In the second experiment ({\em Case~2}), we extend the difficulty by introducing an additional mismatch in the initial observation model. Specifically, all models are initialized with an observation matrix $\mathbf{H}$ rotated by $20^{\circ}$ relative to the true matrix. This simulates a practical scenario where the sensor orientation is imperfectly calibrated at the beginning of operation, in addition to the sudden process noise change at the change point. As shown in Fig.~\ref{fig:partial_info_case2},  degradation is observed for all methods, but  \ac{name} maintains superior robustness. At $0$~dB,  \ac{name} achieves $-9.03$~dB, significantly outperforming KalmanNet ($13.72$~dB) , Unsupervised KalmanNet ($-2.99$~dB), and partial information KF ($-7.32$~dB). Under the more challenging low-SNR setting of $-10$~dB, \ac{name} attains $10.33$~dB, remaining better than KalmanNet ($31.17$~dB) and Unsupervised KalmanNet ($15.70$~dB).

Overall, the results indicate that \ac{name} maintains a consistent advantage over other hybrid-driven methods in both scenarios, demonstrating its ability to handle abrupt parameter changes without explicit \ac{cpd} or parameter reinitialization. The quantitative comparisons confirm that \ac{name} not only narrows the gap with full-information classical methods but, in certain configurations, even surpasses them, highlighting its adaptability to practical situations where both parameter drift and model mismatch occur simultaneously.

\subsection{Non-Linear State Space Model : Lorenz Attractor}
\subsubsection{Setup}
Next, we consider non-linear and chaotic dynamics, using the Lorenz attractor as a benchmark. The Lorenz attractor is a three-dimensional system derived from a simplified model of atmospheric convection, described by a set of continuous-time ordinary differential equations. We discretize the continuous-time process to form an \ac{ss} model. The state evolution is approximated using a Taylor series expansion of the matrix exponential, which is given by:
$$ \myMat{F}(\mathbf{x}_{t}) \approx \myMat{I} + \sum_{j=1}^{J}\frac{(\myMat{A}(\mathbf{x}_{t})\cdot\Delta t)^{j}}{j!}, $$
where $\myMat{A}(\mathbf{x}_{t})$ is the state-dependent dynamics matrix (see \cite{revach2022kalmannet}), $\Delta t$ is the sampling interval, and $J$ is the order of the Taylor series approximation. For our experiments, we set the approximation order $J=5$ and the sampling interval $\Delta t=0.02$s, consistent with the setup for generating the ground truth trajectory.

As the \ac{ss} model is non-linear, we do not use the \ac{kf} and the EM-KF here. Instead, we compare our proposed \ac{name} against several established filters for non-linear dynamics: EKF, UKF, the Particle Filter (PF), as well as the original KalmanNet and an unsupervised variant. All experiments are conducted with a process noise variance of $v=-20$dB, with $\mathbf{Q}= v^2 \mathbf{I}$.

\subsubsection{Full information results}

In the full information scenario, we assume that all filtering algorithms have access to the accurate state-space model, including the correct state-evolution function. The observations are noisy measurements of the true state, where the observation function $h(\cdot)$ is an identity mapping. The performance is evaluated in terms of \ac{mse} and its standard deviation (STD) across various observation noise levels, using the same $1/r^2$ notation defined above.

The results are summarized in Table~\ref{tab:lorenz_full}. At lower SNRs, such as at -10 dB, our proposed method demonstrates competitive performance with an MSE of 18.41 dB. As the SNR increases, the EKF, leveraging its precise model knowledge, achieves the best performance, reaching an MSE of -23.19 dB at an SNR of 20 dB. Our method also shows strong performance in this high-SNR regime with an MSE of -13.89 dB. Notably, our approach cachieves lower MSE than the UKF, KalmanNet, Unsupervised, and PF methods across all tested noise levels, indicating its effectiveness even with perfect model knowledge.

\begin{table*}
\centering
\caption{MSE [dB] and STD for the Lorenz attractor with full information and noisy state observations.}
\label{tab:lorenz_full}
\begin{tabular}{l|cc|cc|cc|cc}
\hline
\multicolumn{1}{c|}{\multirow{2}{*}{\textbf{Method}}} & \multicolumn{2}{c|}{\textbf{-10 dB}} & \multicolumn{2}{c|}{\textbf{0 dB}} & \multicolumn{2}{c|}{\textbf{10 dB}} & \multicolumn{2}{c}{\textbf{20 dB}} \\
\multicolumn{1}{c|}{} & \textbf{MSE} & \textbf{STD} & \textbf{MSE} & \textbf{STD} & \textbf{MSE} & \textbf{STD} & \textbf{MSE} & \textbf{STD} \\ \hline
EKF & 6.63 & 0.52 & -1.43 & 0.43 & -13.47 & 0.43 & -23.19 & 0.54 \\
UKF & 41.72 & 10.41 & 15.99 & 9.08 & 45.97 & 10.41 & 9.01 & 8.94 \\
\textbf{CASA KalmanNet} & \textbf{18.41} & \textbf{4.39} & \textbf{5.83} & \textbf{3.83} & \textbf{-5.20} & \textbf{3.15} & \textbf{-13.89} & \textbf{3.19} \\
KalmanNet & 28.39 & 2.52 & 16.89 & 2.82 & 5.66 & 2.18 & -5.62 & 1.21 \\
Unsupervised & 22.54 & 3.66 & 9.45 & 2.85 & -1.53 & 2.31 & -10.71 & 2.86 \\
PF & 37.49 & 5.05 & 23.05 & 1.75 & 23.66 & 0.97 & 25.58 & 0.24 \\ \hline
\end{tabular}
\end{table*}

\begin{figure}
\centering
\includegraphics[width=0.85\linewidth]{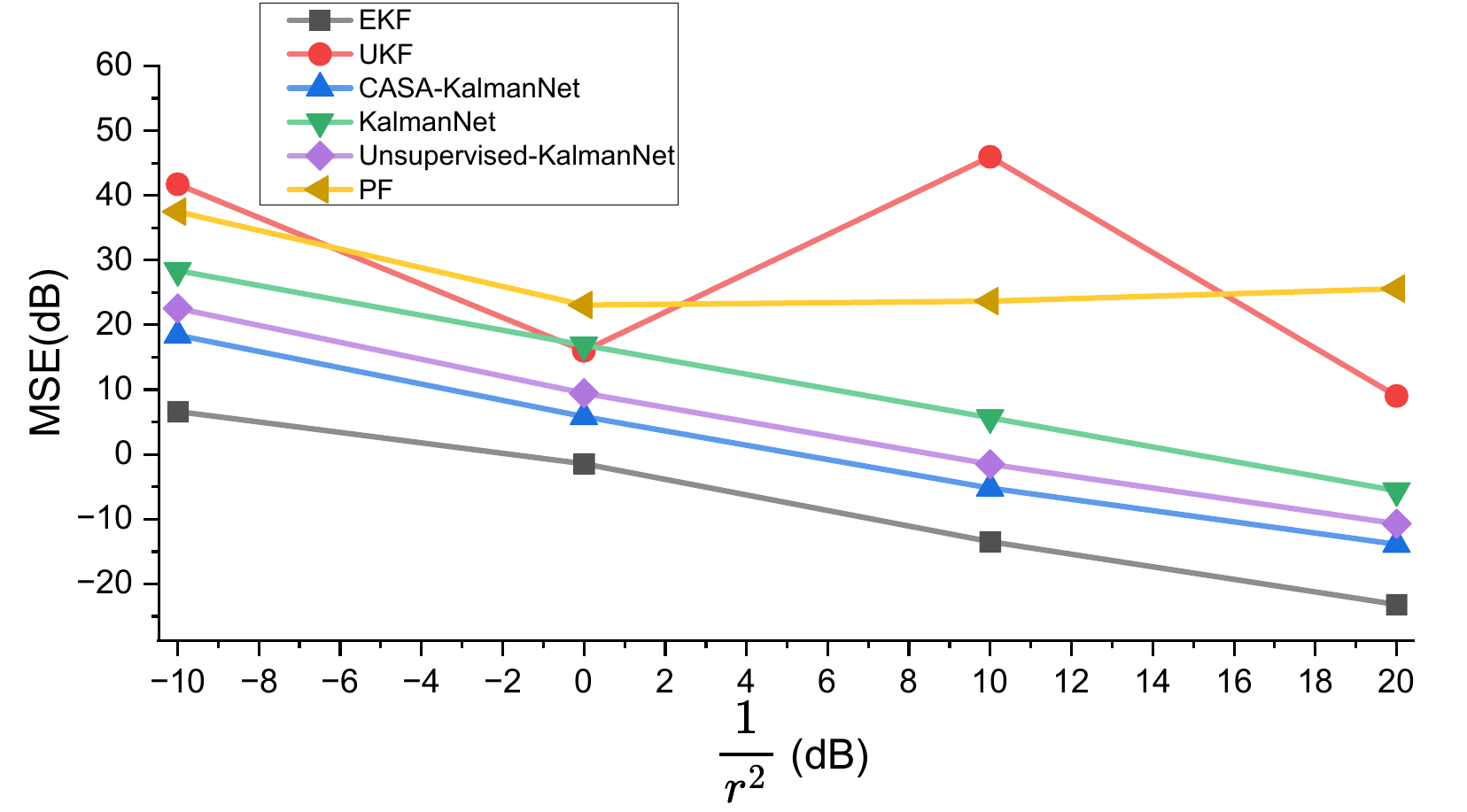}
\caption{Performance comparison on the Lorenz attractor under the full information scenario.}
\label{fig:lorenz_full}
\end{figure}

\subsubsection{Partial information results}

To simulate more realistic conditions, we evaluate the robustness of the filters to model mismatch. In these partial information scenarios, the filters do not have perfect knowledge of the system dynamics. We consider two distinct cases of model uncertainty.

{\em Case 1: Correct initial parameters, unknown dynamics Change.}
In this setup, filters are initialized with the correct model parameters but are unaware of a subsequent change in the system dynamics (e.g., a change point in the process noise covariance $\myMat{Q}$). This tests the adaptability of each filter to unforeseen variations.
Table~\ref{tab:lorenz_partial_correct} presents the results. Under this model mismatch condition, our proposed method demonstrates superior performance and robustness, achieving the lowest MSE across all tested SNRs. For example, at an SNR of 20 dB, our method achieves an MSE of -14.18 dB. This marks a considerable improvement over the EKF (-5.91dB) and KalmanNet (-5.08dB), whose performance degrades due to limited adaptation mechanisms. The  UKF and PF are also affected, reflecting their sensitivity to model inaccuracies.

\begin{table*}
\centering
\caption{MSE [dB] and STD under partial information with correct initial parameters.}
\label{tab:lorenz_partial_correct}
\begin{tabular}{l|cc|cc|cc|cc}
\hline
\multicolumn{1}{c|}{\multirow{2}{*}{\textbf{Method}}} & \multicolumn{2}{c|}{\textbf{-7 dB}} & \multicolumn{2}{c|}{\textbf{0 dB}} & \multicolumn{2}{c|}{\textbf{10 dB}} & \multicolumn{2}{c}{\textbf{20 dB}} \\
\multicolumn{1}{c|}{} & \textbf{MSE} & \textbf{STD} & \textbf{MSE} & \textbf{STD} & \textbf{MSE} & \textbf{STD} & \textbf{MSE} & \textbf{STD} \\ \hline
EKF & 23.11 & 2.57 & 12.85 & 1.61 & 4.60 & 1.06 & -5.91 & 1.74 \\
UKF & 17.98 & 3.15 & 16.15 & 5.16 & 15.93 & 3.91 & 8.77 & 2.96 \\
\textbf{CASA KalmanNet} & \textbf{12.41} & \textbf{3.41} & \textbf{4.02} & \textbf{3.04} & \textbf{-3.75} & \textbf{2.75} & \textbf{-14.18} & \textbf{3.14} \\
KalmanNet & 24.08 & 2.58 & 15.24 & 2.46 & 10.27 & 2.12 & -5.08 & 1.87 \\
Unsupervised & 16.07 & 3.16 & 7.28 & 2.10 & -0.68 & 2.11 & -11.05 & 2.44 \\
PF & 16.07 & 3.16 & 23.94 & 1.40 & 25.37 & 0.51 & 25.64 & 0.12 \\ \hline
\end{tabular}
\end{table*}

\begin{table*}
\centering
\caption{MSE [dB] and STD under partial information with incorrect initial parameters.}
\label{tab:lorenz_partial_incorrect}
\begin{tabular}{l|cc|cc|cc|cc}
\hline
\multicolumn{1}{c|}{\multirow{2}{*}{\textbf{Method}}} & \multicolumn{2}{c|}{\textbf{-10 dB}} & \multicolumn{2}{c|}{\textbf{0 dB}} & \multicolumn{2}{c|}{\textbf{10 dB}} & \multicolumn{2}{c}{\textbf{20 dB}} \\
\multicolumn{1}{c|}{} & \textbf{MSE} & \textbf{STD} & \textbf{MSE} & \textbf{STD} & \textbf{MSE} & \textbf{STD} & \textbf{MSE} & \textbf{STD} \\ \hline
EKF & 28.87 & 2.14 & 12.07 & 1.04 & 2.54 & 1.36 & -5.20 & 1.51 \\
\textbf{CASA KalmanNet} & \textbf{20.19} & \textbf{3.62} & \textbf{3.35} & \textbf{2.32} & \textbf{-3.72} & \textbf{2.02} & \textbf{-7.31} & \textbf{0.97} \\
KalmanNet & 35.44 & 5.38 & 15.47 & 3.16 & 3.08 & 1.45 & -3.87 & 3.87 \\
Unsupervised & 23.77 & 3.11 & 6.42 & 1.78 & -1.85 & 2.10 & -6.94 & 1.59 \\
PF & 44.11 & 5.67 & 23.08 & 1.57 & 24.22 & 0.53 & 25.60 & 0.16 \\ \hline
\end{tabular}
\end{table*}

{\em Case 2: Incorrect initial parameters and unknown dynamics change.}
This case presents a more challenging scenario where filters are initialized with incorrect model parameters from the start and remain unaware of the true system dynamics throughout the estimation process. This evaluates the filters' performance under severe and persistent model mismatch.

The results for this highly mismatched scenario are detailed in Table~\ref{tab:lorenz_partial_incorrect}. Our method continues to exhibit remarkable robustness, achieving the lowest MSE in all conditions, such as -7.31 dB at an SNR of 20 dB. In this setup,  the UKF failed to produce stable estimates and often diverged (hence it is not reported), which highlights the severity of the model mismatch for conventional methods for non-linear \ac{ss} models. In contrast, our \ac{name} not only remains stable but also decisively outperforms all other competing methods, including the EKF (-5.20 dB) and KalmanNet (-3.87 dB), proving its resilience in challenging, ill-specified environments.

The present experiments focus on controlled single-change scenarios with finite sequence length, which allows us to isolate the response of CPDNet-guided adaptation to a clearly defined state-space model mismatch. The proposed mechanism itself is applied at every time step and is therefore not inherently limited to a single change point. In principle, it can respond to multiple abrupt changes when they induce detectable degradation in the filtering behavior. Nevertheless, repeated changes and longer sequences may introduce additional error-accumulation effects, and a systematic evaluation under such conditions is left for future work.

\begin{figure}
\centering
\includegraphics[width=0.85\linewidth]{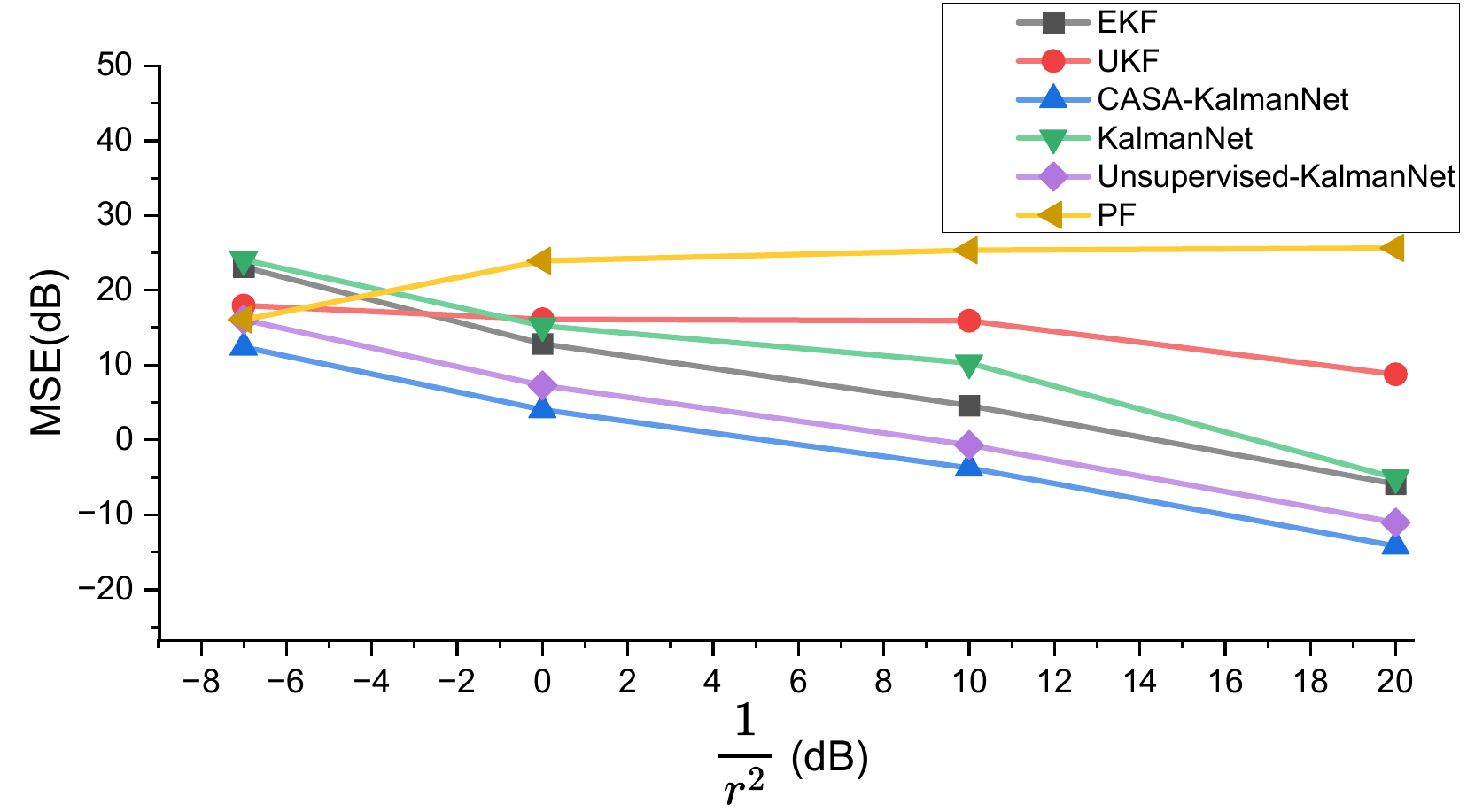}
\caption{Performance comparison on the Lorenz attractor under partial information scenarios with exact initial parameters.}
\label{fig:lorenz_partial_correct_init}
\end{figure}

\begin{figure}
\centering
\includegraphics[width=0.85\linewidth]{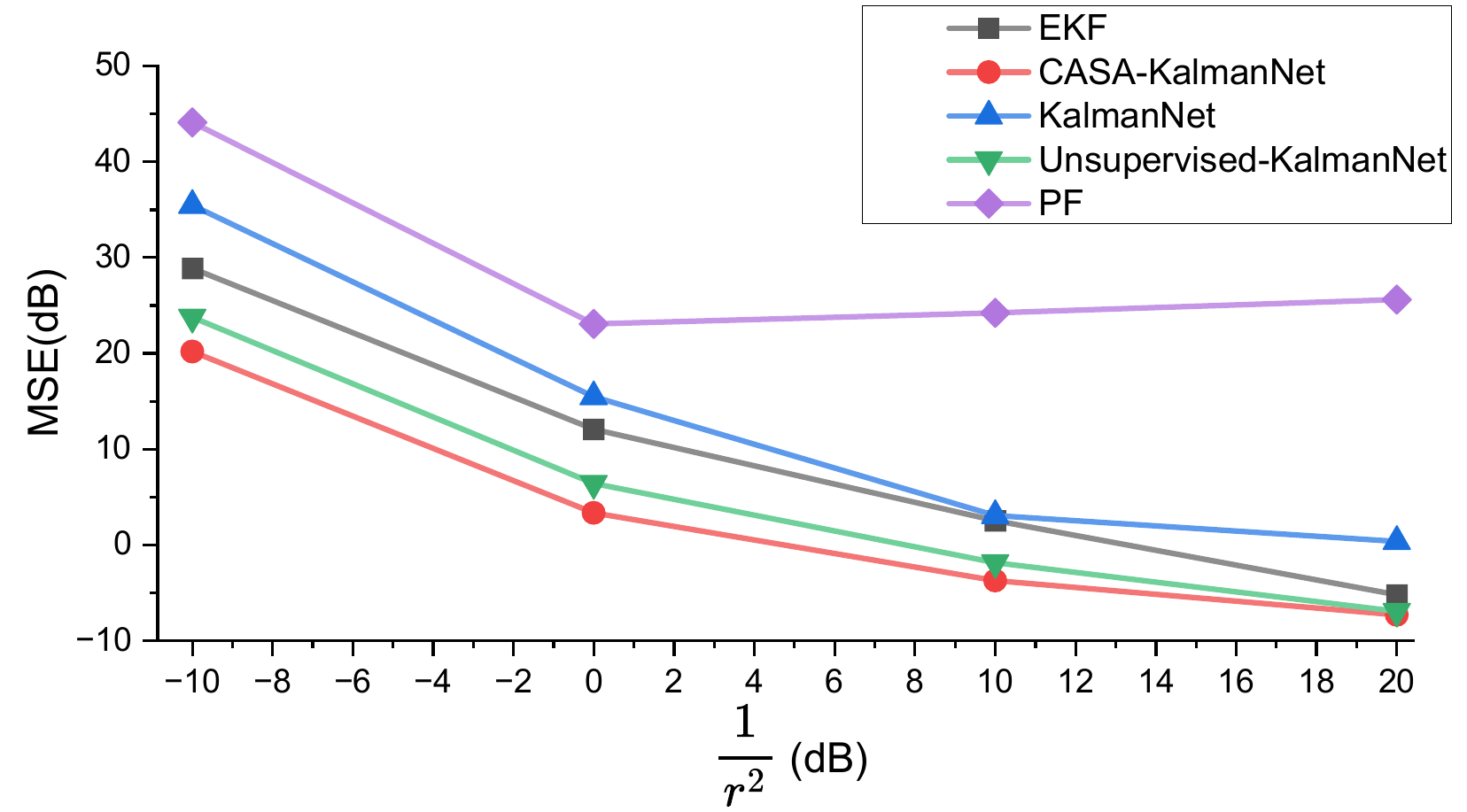}
\caption{Performance comparison on the Lorenz attractor under partial information scenarios with inaccurate initial parameters.}
\label{fig:lorenz_partial_incorrect_init}
\end{figure}

\subsection{Real-World Ground Robot Trajectory: NCLT}
\label{subsec:nclt_gt_r_abrupt}

We further evaluate \ac{name} on a controlled-change experiment constructed from the NCLT ground-robot dataset\cite{ncarlevaris-2015a}. The NCLT ground-truth trajectory is used as the real motion source, and the filtering state is defined as
$\mathbf{x}_t=[p_x,p_y,v_x,v_y]^\top$, where velocity is obtained from smoothed finite differences of the ground-truth position. The observation is a synthetic noisy position measurement,
$\mathbf{y}_t=[p_x,p_y]^\top+\mathbf{v}_t$, under a 2D constant-velocity model. To evaluate label-free adaptation under controlled nonstationarity, the observation-noise covariance $\myMat{R}_t$ is abruptly increased at the midpoint of each test window from
$=0.01 \myMat{I}_2$ to $5.0 \myMat{I}_2$, corresponding to a $500\times$ covariance increase. All compared methods are unaware of this change, and no test-time ground-truth state labels are used for online adaptation.

\begin{table}
\centering
\caption{Controlled $\myMat{R}$-abrupt state estimation results on the NCLT ground-truth trajectory. The state is $[p_x,p_y,v_x,v_y]$, and the observation is synthetic noisy position. An abrupt observation-noise  increase is introduced at the midpoint of each test window.}
\label{tab:nclt_gt_r_abrupt_results}
\small
\setlength{\tabcolsep}{6pt}
\renewcommand{\arraystretch}{1.2}
\begin{tabular}{l|r|r}
\hline
\textbf{Method} & \textbf{MSE(dB)} & \textbf{Updates} \\
\hline
KF & $-4.15$ & $0$ \\

KalmanNet & $9.12$ & $0$ \\

Unsupervised-KalmanNet & $-1.12$ & $22200$ \\

\textbf{CASA-KalmanNet} & $\mathbf{-10.36}$ & $10705$ \\
\hline
\end{tabular}
\end{table}

Table~\ref{tab:nclt_gt_r_abrupt_results} presents the results of the controlled observation-noise abrupt-change experiment driven by the NCLT ground-robot trajectory.
Since the nominal KF has no access to the unannounced covariance change and continues to use the pre-change observation-noise statistics, its performance is degraded under the resulting model mismatch.
The degradation is more pronounced for frozen KalmanNet, whose learned filtering rule is trained under the source-domain noise statistics but deployed in a shifted test-time parameter regime, leading to a substantially larger MSE of $9.12$~dB.
Although continuously updated Unsupervised KalmanNet performs label-free online adaptation, it applies updates indiscriminately at every time step, without assessing whether the current observation is beneficial for adaptation under the increased-noise regime.
Consequently, it incurs over $2\cdot 10^4$ online updates and still underperforms the nominal KF, indicating that continuous unsupervised adaptation may introduce unnecessary parameter drift and substantial adaptation overhead.
In contrast, \ac{name} exploits CPDNet to monitor reliability degradation and selectively regulate online adaptation.
This change-aware update strategy achieves the lowest MSE of $-10.36$~dB, while reducing the number of updates by $51.8\%$  relative to continuous unsupervised updating.
These results demonstrate that \ac{name} provides more robust state estimation under unknown observation-noise nonstationarity while requiring substantially fewer online parameter updates.

\section{Conclusion}
\label{sec:conclusion}
We presented an online adaptation framework for AI-aided Kalman filtering in nonstationary dynamical systems. We introduced CPDNet, a neural change-point detection module that monitors internal features of KalmanNet to provide indicators of reliability degradation, and integrated it into \ac{name}, a self-adaptive architecture that adjusts its online adaptation learning effort based on the severity of detected shifts. This design enables real-time, data-efficient, and autonomous adaptation without requiring additional state labels from the changed regime, while preserving the interpretability and structure of classical filters. Extensive experiments on both linear Gaussian and nonlinear \ac{ss} models demonstrated that \ac{name} consistently outperforms existing learning-based methods under model mismatch and approaches  optimal model-based filters.

\bibliographystyle{IEEEtran}
\bibliography{IEEEabrv,refs}

@STRING{IEEE_J_AES        = "{IEEE} Trans. Aerosp. Electron. Syst."}

@STRING{IEEE_J_VT         = "{IEEE} Trans. Veh. Technol."}

@STRING{IEEE_J_AC         = "{IEEE} Trans. Autom. Control"}

@STRING{IEEE_J_SP         = "{IEEE} Trans. Signal Process."}

@STRING{IEEE_J_WCOM       = "{IEEE} Trans. Wireless Commun."}

@STRING{IEEE_J_KDE        = "{IEEE} Trans. Knowl. Data Eng."}

@STRING{IEEE_J_PROC       = "Proc. {IEEE}"}

@STRING{IEEE_M_SP         = "{IEEE} Signal Process. Mag."}

@inproceedings{Cohen2026EMKalmanNet,
  author    = {Ori Cohen and Xiaoyong Ni and Nir Shlezinger and Tirza Routtenberg},
  title     = {{EM-K}almanNet: {AI}-Aided Kalman Tracking in Partially Known Time-Varying State-Space Models},
  booktitle = {IEEE Sensor Array and Multichannel Signal Processing Workshop (SAM)},
  year      = {2026},
}

@article{masti2021learning,
  author = {Masti, D. and Bemporad, A.},
  title = {Learning nonlinear state–space models using autoencoders},
  volume = {129},
  year = {2021},
  pages = {5–1098},
  language = {it},
  journal = {Automatica}
}

@article{julier2004unscented,
  title={Unscented filtering and nonlinear estimation},
  author={Julier, Simon J and Uhlmann, Jeffrey K},
  journal=IEEE_J_PROC,
  volume={92},
  number={3},
  pages={401--422},
  year={2004},
  publisher={IEEE}
}

@inproceedings{dahan2024uncertainty,
  title={Uncertainty quantification in deep learning based {K}alman filters},
  author={Dahan, Yehonatan and Revach, Guy and Dunik, Jindrich and Shlezinger, Nir},
  booktitle={Proc. IEEE Int. Conf. Acoust. Speech Signal Process.},
  pages={13121--13125},
  year={2024}
}

@article{mehra1972approaches,
  title={Approaches to adaptive filtering},
  author={Mehra, Raman},
  journal=IEEE_J_AC,
  volume={17},
  number={5},
  pages={693--698},
  year={1972},
  publisher={IEEE}
}

@article{chen2025maml,
  title={{MAML-KalmanNet}: A Neural Network-Assisted {K}alman Filter Based on Model-Agnostic Meta-Learning},
  author={Chen, Shanli and Zheng, Yunfei and Lin, Dongyuan and Cai, Peng and Xiao, Yingying and Wang, Shiyuan},
  journal=IEEE_J_SP,
  volume={73}, 
  pages={988--1003},
  year={2025},
  publisher={IEEE}
}

@article{schmidt1981kalman,
  title={The {K}alman filter-Its recognition and development for aerospace applications},
  author={Schmidt, Stanley F},
  journal={J. Guid. Control Dyn.},
  volume={4},
  number={1},
  pages={4--7},
  year={1981}
}

@article{kalman1960new,
title={A new approach to linear filtering and prediction problems},
	author={Kalman, Rudolph Emil},
	journal = {J. Basic Eng.},
	volume = {82},
	number = {1},
	pages = {35-45},
	year={1960}
}

@article{liu2021towards,
  title={Towards out-of-distribution generalization: A survey},
  author={Liu, Jiashuo and Shen, Zheyan and He, Yue and Zhang, Xingxuan and Xu, Renzhe and Yu, Han and Cui, Peng},
  journal={arXiv preprint arXiv:2108.13624},
  year={2021}
}

@article{gannot2008kalman,
  title={The {K}alman Filter},
  author={Gannot, Sharon and Yeredor, Arie},
  journal={Springer Handbook of Speech Processing},
  pages={135--160},
  year={2008},
  publisher={Springer}
}

@article{shlezinger2020model,
	title={Model-Based Deep Learning},
	author={Shlezinger, Nir and Whang, Jay and Eldar, Yonina C and Dimakis, Alexandros G},
	journal=IEEE_J_PROC,
	volume={111},
	number={5},
	pages={465--499},
	year={2023}
}

@article{shlezinger2022discriminative,
  title={Discriminative and Generative Learning for Linear Estimation of Random Signals [Lecture Notes]},
  author={Shlezinger, Nir and Routtenberg, Tirza},
  journal=IEEE_M_SP,
  volume={40},
  number={6},
  pages={75--82},
  year={2023}
}

@book{durbin2012time,
  title={Time Series Analysis by State Space Methods},
  author={Durbin, James and Koopman, Siem Jan},
  year={2012},
  publisher={Oxford University Press}
}

@article{revach2022kalmannet,
  title={Kalman{N}et: Neural network aided {K}alman filtering for partially known dynamics},
  author={Revach, Guy and Shlezinger, Nir and Ni, Xiaoyong and Escoriza, Adria Lopez and Van Sloun, Ruud JG and Eldar, Yonina C},
  journal=IEEE_J_SP,
  volume={70},
  pages={1532--1547},
  year={2022},
  publisher={IEEE}
}

@article{buchnik2024gsp,
  title={{GSP-KalmanNet}: Tracking graph signals via neural-aided {K}alman filtering},
  author={Buchnik, Itay and Sagi, Guy and Leinwand, Nimrod and Loya, Yuval and Shlezinger, Nir and Routtenberg, Tirza},
  journal=IEEE_J_SP,
  volume={72},
  pages={3700--3716},
  year={2024},
  publisher={IEEE}
}

@article{buchnik2023latent,
  title={Latent-{K}alman{N}et: Learned {K}alman filtering for tracking from high-dimensional signals},
  author={Buchnik, Itay and Revach, Guy and Steger, Damiano and Van Sloun, Ruud JG and Routtenberg, Tirza and Shlezinger, Nir},
  journal=IEEE_J_SP,
  volume={72},
  pages={352--367},
  year={2023},
  publisher={IEEE}
}

@article{wang2024nonlinear,
  title={Nonlinear {K}alman Filtering based on Self-Attention Mechanism and Lattice Trajectory Piecewise Linear Approximation},
  author={Wang, Jiaming and Geng, Xinyu and Xu, Jun},
  journal={arXiv preprint arXiv:2404.03915},
  year={2024}
}

@inproceedings{satorras2019combining,
title={Combining Generative and Discriminative Models for Hybrid Inference},
author={Satorras, Victor Garcia and Akata, Zeynep and Welling, Max},
booktitle={Proc. Adv. Neural Inf. Process.},
pages={13802--13812},
year={2019}
}

@article{lan2023variational,
  title={Variational nonlinear {K}alman filtering with unknown process noise covariance},
  author={Lan, Hua and Hu, Jinjie and Wang, Zengfu and Cheng, Qiang},
  journal=IEEE_J_AES,
  volume={59},
  number={6},
  pages={9177--9190},
  year={2023},
  publisher={IEEE}
}

@inproceedings{ni2024adaptive,
  title={Adaptive {K}alman{N}et: Data-driven {K}alman filter with fast adaptation},
  author={Ni, Xiaoyong and Revach, Guy and Shlezinger, Nir},
  booktitle={Proc. IEEE Int. Conf. Acoust. Speech Signal Process.},
  pages={5970--5974},
  year={2024}
}

@article{arasaratnam2009cubature,
  title={Cubature {K}alman filters},
  author={Arasaratnam, Ienkaran and Haykin, Simon},
  journal=IEEE_J_AC,
  volume={54},
  number={6},
  pages={1254--1269},
  year={2009},
  publisher={IEEE}
}

@article{ni2022rtsnet,
  title={{RTSNet}: Learning to Smooth in Partially Known State-Space Models},
  author={Revach, Guy and Ni, Xiaoyong and Shlezinger, Nir and van Sloun, Ruud JG and Eldar, Yonina C},
  journal=IEEE_J_SP,
  volume={71},
  pages={4441--4456},
  year={2023}
}

@inproceedings{revach2022unsupervised,
  title={Unsupervised learned {K}alman filtering},
  author={Revach, Guy and Shlezinger, Nir and Locher, Timur and Ni, Xiaoyong and van Sloun, Ruud JG and Eldar, Yonina C},
  booktitle={Proc. Eur. Signal Process. Conf.},
  pages={1571--1575},
  year={2022},
  organization={IEEE}
}

@article{ghosh2023danse,
  title={{DANSE}: Data-driven Non-linear State Estimation of Model-free Process in Unsupervised Learning Setup},
  author={Ghosh, Anubhab and Honor{\'e}, Antoine and Chatterjee, Saikat},
  journal=IEEE_J_SP,
  volume={72},
  pages={1824--1838},
  year={2024},
  publisher={IEEE}
}

@article{choi2023split,
  title={Split-{KalmanNet}: A robust model-based deep learning approach for state estimation},
  author={Choi, Geon and Park, Jeonghun and Shlezinger, Nir and Eldar, Yonina C and Lee, Namyoon},
  journal=IEEE_J_VT,
  volume={72},
  number={9},
  pages={12326--12331},
  year={2023},
  publisher={IEEE}
}

@inproceedings{gama2004learning,
  title={Learning with drift detection},
  author={Gama, Joao and Medas, Pedro and Castillo, Gladys and Rodrigues, Pedro},
  booktitle={Brazilian Symposium on Artificial Intelligence},
  pages={286--295},
  year={2004},
  organization={Springer}
}

@article{gruslys2016memory,
  title={Memory-efficient backpropagation through time},
  author={Gruslys, Audrunas and Munos, R{\'e}mi and Danihelka, Ivo and Lanctot, Marc and Graves, Alex},
  journal={Proc. Adv. Neural Inf. Process.},
  volume={29},
  year={2016}
}

@article{lu2018learning,
  title={Learning under concept drift: A review},
  author={Lu, Jie and Liu, Anjin and Dong, Fan and Gu, Feng and Gama, Joao and Zhang, Guangquan},
  journal=IEEE_J_KDE,
  volume={31},
  number={12},
  pages={2346--2363},
  year={2018},
  publisher={IEEE}
}

@article{Shumway1982,
    title = {An Approach to Time Series Smoothing and Forecasting Using the {EM} Algorithm},
    author = {Shumway, R. H. and Stoffer, D. S.},
    year = {1982},
    journal = {J. Time Ser. Anal.},
    volume = {3},
    number = {4},
    pages = {253-264},
}

@article{ghosh2024deepbayes,
  title={Deep{B}ayes—An estimator for parameter estimation in stochastic nonlinear dynamical models},
  author={Ghosh, Anubhab and Abdalmoaty, Mohamed and Chatterjee, Saikat and Hjalmarsson, H{\aa}kan},
  journal={Automatica},
  volume={159},
  pages={111327},
  year={2024},
  publisher={Elsevier}
}

@inproceedings{jouaber2021nnakf,
  title={{NNAKF}: A neural network adapted {K}alman filter for target tracking},
  author={Jouaber, Sami and Bonnabel, Silvere and Velasco-Forero, Santiago and Pilte, Marion},
  booktitle={Proc. IEEE Int. Conf. Acoust. Speech Signal Process.},
  pages={4075--4079},
  year={2021}
}

@article{xu2024ekfnet,
  title={{EKFNet}: Learning system noise covariance parameters for nonlinear tracking},
  author={Xu, Liang and Niu, Ruixin},
  journal=IEEE_J_SP,
  volume={72},
  pages={3139--3152},
  year={2024},
  publisher={IEEE}
}

@inproceedings{welling2021hkf,
  title={Neural Augmentation of {K}alman Filter with Hypernetwork for Channel Tracking},
  author={Pratik, Kumar and Amjad, Rana Ali and Behboodi, Arash and Soriaga, Joseph B and Welling, Max},
  booktitle={Proc. IEEE Glob. Commun. Conf.},
  year={2021}
}

@inproceedings{uzlaner2024concept,
  title={Concept drift detection for deep learning aided receivers in dynamic channels},
  author={Uzlaner, Nicole and Raviv, Tomer and Shlezinger, Nir and Todros, Koby},
  booktitle={Proc. IEEE Workshop Signal Process. Adv. Wirel.},
  pages={371--375},
  year={2024}
}

@article{aminikhanghahi2017survey,
  title={A survey of methods for time series change point detection},
  author={Aminikhanghahi, Samaneh and Cook, Diane J},
  journal={Knowl. Inf. Syst.},
  volume={51},
  number={2},
  pages={339--367},
  year={2017},
  publisher={Springer}
}

@article{uzlaner2025asynchronous,
  title={Asynchronous Online Adaptation via Modular Drift Detection for Deep Receivers},
  author={Uzlaner, Nicole and Raviv, Tomer and Shlezinger, Nir and Todros, Koby},
  journal=IEEE_J_WCOM,
  volume={24},
  number={5},
  pages={4454-4468},
  year={2025},
  publisher={IEEE}
}

@article{shlezinger2025artificial,
  title={Artificial Intelligence-Aided {K}alman Filters: {AI}-Augmented Designs for {K}alman-Type Algorithms},
  author={Shlezinger, Nir and Revach, Guy and Ghosh, Anubhab and Chatterjee, Saikat and Tang, Shuo and Imbiriba, Tales and Dunik, Jindrich and Straka, Ondrej and Closas, Pau and Eldar, Yonina C},
  journal=IEEE_M_SP,
  year={2025, early access},
  publisher={IEEE}
}

@article{jung2025state,
title={State Estimation with 1-Bit Observations and Imperfect Models: Bussgang Meets {K}alman in Neural Networks}, 
author={Chaehyun Jung and TaeJun Ha and Hyeonuk Kim and Jeonghun Park},
      year={2025},
  journal={arXiv preprint arXiv:2507.17284}
}

@phdthesis{sutskever2013training,
  title={Training recurrent neural networks},
  author={Sutskever, Ilya},
  year={2013}
}

@article{miller2020adversarial,
  title={Adversarial learning targeting deep neural network classification: A comprehensive review of defenses against attacks},
  author={Miller, David J and Xiang, Zhen and Kesidis, George},
  journal=IEEE_J_PROC,
  volume={108},
  number={3},
  pages={402--433},
  year={2020},
  publisher={IEEE}
}

@article{shlezinger2023model,
  title={Model-Based Deep Learning},
  author={Shlezinger, Nir and Eldar, Yonina C},
  journal={Found. Trends Signal Process.},
  volume={17},
  number={4},
  pages={291--416},
  year={2023},
  publisher={Now Publishers, Inc.}
}

@article{imbiriba2023augmented,
  author={Imbiriba, Tales and Straka, Ondřej and Duník, Jindřich and Closas, Pau},
  journal=IEEE_J_AES, 
  title={Augmented Physics-Based Machine Learning for Navigation and Tracking}, 
  year={2024},
  volume={60},
  number={3},
  pages={2692-2704}
}

@book{bar2004estimation,
  title={Estimation With Applications to Tracking and Navigation: Theory Algorithms and Software},
  author={Bar-Shalom, Yaakov and Li, X Rong and Kirubarajan, Thiagalingam},
  year={2004},
  publisher={John Wiley \& Sons}
}

@article{adams2007bayesian,
  title={Bayesian online change point detection},
  author={Adams, Ryan Prescott and MacKay, David JC},
  journal={arXiv preprint arXiv:0710.3742},
  year={2007}
}

@inproceedings{alami2020restarted,
  title={Restarted {B}ayesian online change-point detector achieves optimal detection delay},
  author={Alami, R{\'e}da and Maillard, Odalric and F{\'e}raud, Raphael},
  booktitle={Proc. Int. Conf. Mach. Learn.},
  pages={211--221},
  year={2020},
  organization={PMLR}
}

@article{harchaoui2008kernel,
  title={Kernel change-point analysis},
  author={Harchaoui, Zaid and Moulines, Eric and Bach, Francis},
  journal={Proc. Adv. Neural Inf. Process.},
  volume={21},
  year={2008}
}

@article{song2024kernal,
  title={Practical and powerful kernel-based change-point detection},
  author={Song, Hoseung and Chen, Hao},
  journal=IEEE_J_SP,
  volume={72}, 
  pages={5174--5186},
  year={2024},
  publisher={IEEE}
}

@ARTICLE{huang2021variational,
  author={Huang, Yulong and Zhang, Yonggang and Shi, Peng and Chambers, Jonathon},
  journal={IEEE Trans. Automat. Contr.}, 
  title={Variational Adaptive {K}alman Filter With {Gaussian-Inverse-Wishart} Mixture Distribution}, 
  year={2021},
  volume={66},
  number={4},
  pages={1786-1793}  }

@ARTICLE{bai2022novel,
  author={Bai, Mingming and Huang, Yulong and Chen, Badong and Zhang, Yonggang},
  journal={IEEE Trans. Syst. Man Cybern. Syst.}, 
  title={A Novel Robust {K}alman Filtering Framework Based on Normal-Skew Mixture Distribution}, 
  year={2022},
  volume={52},
  number={11},
  pages={6789-6805}
  }

@ARTICLE{he2025distributed,
  author={He, Jiacheng and Wang, Gang and Feng, Zhenyu and Zhong, Shan and Zhang, Ping and Peng, Bei},
  journal={IEEE Trans. Instrum. Meas.}, 
  title={Distributed Generalized Minimum Error Entropy Unscented {K}alman Filter Under Hybrid Attacks Without Prior Knowledge}, 
  year={2025},
  volume={74},
  number={},
  pages={1-12}  }

@article{COHEN2025110221,
title = {Adaptive {K}alman-Informed Transformer},
journal = {Eng. Appl. Artif. Intell.},
volume = {146},
pages = {110221},
year = {2025},
issn = {0952-1976},
author = {Nadav Cohen and Itzik Klein}
}

@article{ncarlevaris-2015a,
  author  = { Nicholas Carlevaris-Bianco and Arash K. Ushani and Ryan M. Eustice },
  title   = { University of {Michigan} North Campus long-term vision and lidar dataset },
  journal = { Int. J. Robot. Res. },
  year    = { 2015 },
  volume  = { 35 },
  number  = { 9 },
  pages   = { 1023--1035 }
}

@article{102704,
  author   = {Gutman, P.-O. and Velger, M.},
  journal  = {IEEE Trans. Aerosp. Electron. Syst.},
  title    = {Tracking targets using adaptive Kalman filtering},
  year     = {1990},
  volume   = {26},
  number   = {5},
  pages    = {691-699}}

\end{document}